\documentclass[fleqn,usenatbib]{mnras}

\usepackage[T1]{fontenc}
\usepackage{ulem} 

\DeclareRobustCommand{\VAN}[3]{#2}
\let\VANthebibliography\thebibliography
\def\thebibliography{\DeclareRobustCommand{\VAN}[3]{##3}\VANthebibliography}

\usepackage{graphicx}	
\usepackage{amsmath}	
\usepackage{amssymb}	

\usepackage{lineno}

\title[Hunting for the candidates of misclassified sources]
{Hunting for the candidates of misclassified sources in LSP BL Lacs using Machine learning }

\author[S. J. Kang et al.]{
Shi-Ju Kang,$^{1}$\thanks{E-mail: kangshiju@alumni.hust.edu.cn (SJK)}
Yong-Gang Zheng,$^{2}$\thanks{E-mail: ynzyg@ynu.edu.cn (YGZ)}
and Qingwen Wu$^{3}$\thanks{E-mail: qwwu@hust.edu.cn (QW)}
\\
$^{1}$School of Physics and Electrical Engineering, Liupanshui Normal University, Liupanshui, Guizhou, 553004, China\\
$^{2}$Department of Physics, Yunnan Normal University, Kunming, Yunnan, 650092, China\\
$^{3}$Department of Astronomy, School of Physics, Huazhong University of Science and Technology, Wuhan, Hubei, 430074, China
}

\date{Accepted 2023 August 09. Received 2023 July 17; in original form 2023 March 21}

\pubyear{2023}

\begin{document}
\label{firstpage}
\pagerange{\pageref{firstpage}--\pageref{lastpage}}

\maketitle

\begin{abstract}
An equivalent width (EW) based classification may cause the erroneous judgement to the flat spectrum radio quasars (FSRQs) and BL Lacerate objects (BL Lac) due to the diluting the line features by dramatic variations in the jet continuum flux. To help address the issue, the present paper explore the possible intrinsic classification on the bias of a random forest supervised machine learning algorithm. In order to do so, we compile a sample of 1680 Fermi blazars that have both gamma-rays and radio-frequencies data available from the 4LAC-DR2 catalog, which includes 1352 training and validation samples and 328 forecast samples. By studying the results for all of the different combinations of 23 characteristic parameters, we found that there are 178 optimal parameters combinations (OPCs) with the highest accuracy ($\simeq$ 98.89\%). Using the combined classification results from the nine combinations of these OPCs to the 328 forecast samples, we predict that there are 113 true BL Lacs (TBLs) and 157 false BL Lacs (FBLs) that are possible intrinsically FSRQs misclassified as BL Lacs. The FBLs show a clear separation from TBLs and FSRQs in  the $\gamma$-ray photon spectral index, $\Gamma_{\rm ph}$, and X-band radio flux, ${\rm log}{F_{R}}$, plot.  Phenomenally, existence a BL Lac to FSRQ (B-to-F) transition zone is suggested, where the FBLs are in the stage of transition from BL Lacs to FSRQs. Comparing the LSP Changing-Look Blazars (CLBs) reported in the literatures, the majority of LSP CLBs are located at the B-to-F zone. We argue that the FBLs located at B-to-F transition zone are the most likely Candidates of CLBs.
\end{abstract}




\section{Introduction}

Blazars are a peculiar sub-class of radio-loud active galactic nucleis (AGNs) with a relativistic jet pointed towards us, whose multi-wavelength spectral energy distributions (SEDs) dominantly originates from the non-thermal emission in the relativistic jet \citep{1995PASP..107..803U}. According to the strength of the optical spectral lines (e.g., equivalent width, EW, of the spectral line is greater or less than  {5\AA}), blazars come in two flavors: flat spectrum radio quasars (FSRQs) with the stronger emission lines (EW {$\ge$ {5\AA}}), and BL Lacerate objects (BL Lacs) that the spectral lines are fainter or even absent (EW < {5\AA}) in their optical spectra \citep{1991ApJ...374..431S,1991ApJS...76..813S}. The broadband SEDs of blazars normally exhibits a two-hump structure in the ${\rm log \nu-log \nu F_{\nu}}$ space. The lower energy bump usually originates from synchrotron radiation generated by non-thermal electrons in the jet, while the second bump mainly originates from inverse Compton (IC) scattering. Based on the peak frequency ($\nu^{\rm S}_{\rm p}$) of the lower energy bump,  blazars are also normally subclassified as low (LSP, e.g., $\nu^{\rm S}_{\rm p}<10^{14}$ Hz), intermediate (ISP, e.g., $10^{14}~\rm Hz<\nu^{\rm S}_{\rm p}<10^{15}$ Hz) and high-synchrotron-peaked (HSP, e.g., $\nu^{\rm S}_{\rm p}>10^{15}$ Hz) blazars (\citealt{2010ApJ...716...30A}; \citealt{2016ApJS..226...20F}). Most HSP and ISP blazars have been classified as BL Lacs,  while the LSP class contains both FSRQs and LSP  BL Lacs \citep{2019Galax...7...20B,2022Galax..10...35P}.

\vspace{2mm} 

The EW-based classification is simple, and provides some clues that these sources with whether intrinsically strong (FSRQ) or weak (BL Lac) emission lines. However, the EW of blazars doesn't show bimodal distribution in observations. The EW-based classification with the EW value of 5\AA~is rather arbitrary, for instance, in rest frame by \cite{1991ApJ...374..431S} or observed-frame by \cite{1991ApJS...76..813S}. Furthermore, the continuum as used in EW is complex in blazars, where the optical emission can be seriously contaminated by the variable jet emission. Hence, the line EW can dramatically vary from one state to another for the same source. So, the EW-based classification may have selection effects (e.g., \citealt{2012MNRAS.420.2899G,2013MNRAS.431.1914G,2015MNRAS.446L..41P}), and may lead to several misclassifications, since the broad lines can be swamped by a particularly strong (and possibly beamed) continuum \citep[e.g.,][]{2014ApJ...797...19R,2019RNAAS...3...92P}.

 \vspace{2mm}
 
For instance,  a blazar with intrinsically very luminous emission lines can temporarily appear as a BL Lac, with very small EW, if its jet flux happens to be more luminous than usual \citep[e.g.,][]{2021ApJ...913..146M}. The EW of the changing-look blazar  (CLBs) B2 1420+32 shows a  floats at 5{\AA} during between FSRQ and BL Lac state transitions. On the contrary,  these transitional objects may show broad lines in the optical band when the continuum is low (e.g., \citealt{2014ApJ...797...19R}), where, during a particularly low state, a BL Lac can show emission lines with EW larger than the 5{\AA} limit (as it happened to BL Lac itself; \citealt{1995ApJ...452L...5V,1996MNRAS.281..737C, 2010A&A...516A..59C}). In addition, the EW-based classification is also affected by the strong non-thermal emission (e.g., \citealt{2011MNRAS.414.2674G}), or a high Doppler boosting/ jet bulk Lorentz factor variability  (e.g., \citealt{2009A&A...496..423B}), and the effect of high redshift, i.e. ${\rm H}\alpha$ line, one of the strongest emission line maybe falls outside the optical window so that it is not detected (e.g., \citealt{2015MNRAS.449.3517D}; \citealt{2014ApJ...794....8S}).

 \vspace{2mm}

The possible physical reasons for the physical difference between FSRQs and BL Lacs has been extensively explored. For instance, the physical difference between FSRQ and BL Lac may be  attributed to different accretion models (e.g.,  \citealt{2002ApJ...570L..13C,2002ApJ...579..554W, 2003ApJ...599..147C,2007AJ....133.2187D, 2009MNRAS.396L.105G, 2009ApJ...694L.107X, 2014MNRAS.445...81S, 2015AJ....150....8C, 2017FrASS...4....6F}; \citealt{2018ApJS..235...39C, 2018MNRAS.473.2639G}; \citealt{2019MNRAS.482L..80B,2019MNRAS.486.3465M,2021MNRAS.505.4726K,2022Galax..10...35P} for more details and reference therein). Which may provide different environment, rich of photons or not (different external seed photons), from outside of the jet \citep[e.g.,][]{1998MNRAS.301..451G,2016Galax...4...36G,2017MNRAS.469..255G,2022Galax..10...35P}, for the different cooling of relativistic electrons (\citealt{1998MNRAS.301..451G}), where FSRQs with a standard cold accretion disk providing  a fast cooling environment and BL Lacs with an advection-dominated accretion flow (ADAF; e.g.,  \citealt{2014ARA&A..52..529Y})  providing  a slow cooling environment. In addition, it may also be attributed to the beaming effect (e.g., \citealt{2003A&A...407..899F}); the spin of a central black hole \citep[e.g., see][]{2016RAA....16...54B,2018MNRAS.473.2639G}; and/or mass accretion rate on to the central black hole (e.g., see \citealt{2019MNRAS.482L..80B});  and/or both the mass accretion rate and magnetic field strength (e.g., see \citealt{2019MNRAS.486.3465M}).

 \vspace{2mm}

Motivated by the observational background, some more physical classifications for blazars  are proposed (e.g., \citealt{2004MNRAS.351...83L}; \citealt{2010ApJ...716...30A}; \citealt{2011MNRAS.414.2674G}; \citealt{2011ApJ...740...98M}; \citealt{2012MNRAS.420.2899G}; \citealt{2016A&A...592A..22H}): for instance, which is based on the sources with intrinsically weak or strong OII and OIII emission lines \citep{2004MNRAS.351...83L}; based on the different accretion rates (the luminosity of the broad line region relative to the Eddington luminosity) of the two subclasses of blazars (e.g., \citealt{2011MNRAS.414.2674G,2012MNRAS.421.1764S} ); based on the ionizing luminosity emitted from the accretion disc (e.g.,  \citealt{2012MNRAS.420.2899G,2013MNRAS.431.1914G,2015MNRAS.450.2404G}); alternatively, based on the kinematic features of their radio jets (e.g., \citealt{2016A&A...592A..22H}), etc.

\vspace{2mm}

The differences between LSP BL Lacs and FSRQs or remaining  BL Lacs (HSP and ISP) have been widely argued/debated by some works (e.g., \citealt{2012ApJ...757...25L}). Some sources classified as LSP BL Lacs have strong emission lines, and are more strongly beamed than the rest of the BL Lac object population (e.g., \citealt{2012ApJ...757...25L}). Alternatively, some other sources classified as FSRQs may have weaker emission lines. The dichotomy between LSP BL Lacs and FSRQs is complicated in the classification of blazars, which may be misclassified. Some of the LSP BL Lacs may not actually be BL Lacs (e.g., \citealt{2009MNRAS.396L.105G}; \citealt{2012MNRAS.420.2899G}). In fact, it is possible that BL Lacertae itself is not actually a BL Lac object \citep{1995ApJ...452L...5V}, when its jet continuum is in a particularly low state, that can show emission lines with EW larger than the 5\AA. Some objects classified as LSP BL Lacs are actually FSRQs with exceptionally strong jet emission overpowering the emission lines (e.g., see \citealt{2012ApJ...757...25L} and references therein). The lack of obvious broad lines leads the astronomical community to misclassify some sources as BL Lac objects. In addition, some authors found that some parameters show a very broad distribution for LSP BL Lacs, which is somewhat bimodal (e.g., \citealt{2019ApJ...879..107F}; \citealt{2022MNRAS.515.2215C}). For example, the jet power of LSP BL Lacs shows a very broad bimodal distribution, which suggests that they may contain two populations, one is actually FSRQs with at high redshifts, others with a lower power located at low redshifts, similar to actual BL Lacs (e.g., \citealt{2019ApJ...879..107F}). In our previous work, we found that the distribution of the peak frequency of the synchrotron radiation, gamma-ray photon spectral index, and the X-band (8.4 GHz) flux density showed a similar bimodal for the LSP subclass; one distribution group similar to the BL Lacs and another similar to the FSRQs (\citealt{2022MNRAS.515.2215C}). We suggested that there are 47 LSP BL Lacs  may be FSRQs. Which may indicate that some LSP-BL Lacs may belong to actually BL Lacs and others are essentially FSRQs, and vice versa.

\vspace{2mm}

Motivated by these observational background,  we employ a random forest supervised machine learning (SML) algorithms, and aim to try to diagnose/evaluate some BL Lacs showing the observational characteristics of FSRQ type sources that could be potential FSRQs using the 4FGL catalog based on the  {more} observational properties. Section 2 introduces the Random Forest supervised machine learning algorithms. The method used to select the parameters and data sample from the catalog is described in Section 3. The optimal combinations of parameters  and classification results are reported in Section 4. A comparison with other results is presented in Section 5. The discussion  and conclusion are shown in Section 6.

\section{Random forest SML algorithms} \label{sec:method}
\vspace{3mm}

Random forests  algorithm is a popular, well established SML algorithms, which  has been widely used in astronomical research (e.g., see \citealt{2019arXiv190407248B}; \citealt{2012msma.book.....F,Kabacoff2015R} for the reviews). The original random forests proposal (\citealt{2001MachL..45....5B}), which has evolved over time, transforms a training sample into a large collection of decisions trees (i.e., forest). These trees are used to conduct an extensive voting scheme, which enhances the classification and the prediction accuracy of the model. The random forests algorithm has numerous advantages, including accuracy, scalability, and the ability to address challenging datasets. In terms of accuracy, the random forests approach has outperformed alternative approaches, for instance, decision trees, support vector machines, etc. (e.g., \citealt{JMLRv15delgado2014a,2019ApJ...872..189K, 2019ApJ...887..134K, 2021RAA....21...15Z}). Random forests successfully builds predictive models for uneven datasets, for example, those with large amounts of missing data or a relatively limited ratio of observations in comparison to the number of variables. The random forests approach also generates out-of-bag error rates, in addition to measures indicating the relative importance of the variables. However, due to the large number of trees (default 500 trees), it is difficult to understand the classification rules and make communications, for instance, sharing the classification rules with others can be extremely challenging.

 \vspace{2mm}

In the last decade, machine learning has been widely applied to the blazars classification
(e.g., \citealt{2014ApJ...782...41D};
\citealt{2014ApJS..215...14D,2019ApJS..242....4D};
\citealt{2016MNRAS.462.3180C,2021JHEAp..29...40C};
\citealt{2017MNRAS.470.1291S};
\citealt{2019ApJ...871...94K,2019ApJ...887...18K};
 \citealt{2019ApJ...872..189K, 2019ApJ...887..134K};
 \citealt{ 2020MNRAS.498.1750A};
\citealt{2020MNRAS.493.1926K};
\citealt{2021MNRAS.505.1268F};
\citealt{2021ApJ...923...75K};
\citealt{2021ApJ...908..177K,2023ApJ...943..167K};
\citealt{2022JCAP...04..023B};
\citealt{2022Univ....8..436F};
\citealt{2023ApJ...946..109A}, 
\citealt{2023MNRAS.519.3000S};
etc). Due to random forest classification algorithm frequently performs well with a higher prediction accuracy (e.g., \citealt{2019ApJ...872..189K, 2019ApJ...887..134K}, etc), it is employed and used in this work.

\vspace{2mm}

Many software packages are available for random forests algorithms. The randomForest package  (\citealt{R_randomForest}) in R\footnote{\url{https://www.R-project.org/}} {(R version 4.1.2, \citealt{R_code} ) } is selected and used to fit a random forest in this work. Additionally, the accuracy of the model is calculated using the $classAgreement()$ function in the e1071 package \citep{R_e1071}. An R package ``snowfall”  is employed to make parallel programming \citep{R_snowfall}.

\section{Sample  and Parameters Preparation} \label{Sample_data_pre}
\vspace{3mm}

\begin{table*}
\centering
\caption{The Results of the Two-sample K-S Test for 651 FSRQs and 701 ISP (and HSP) BL Lacs}
\begin{tabular}{ccccccccc}
\hline \hline 
Label&    &Selected Parameters  &&D  of K-S test&&  $p$   of K-S test&&  MeanDecreaseGini    \\ 
(1)    &     & (2) &     &        (3) &      &         (4) &     &        (5)      \\ 
\hline 
1	&	&	$	{\rm PL\_Index}         $	&	&	0.845 	&	&	$<$1E-16	&	&	82.172 	\\
2	&	&	$	{\rm X\_band}           $	&	&	0.816 	&	&	$<$1E-16	&	&	64.389 	\\
3	&	&	$	{\rm Pivot\_Energy}    	$	&	&	0.805 	&	&	$<$1E-16	&	&	58.089 	\\
4	&	&	$	{\rm HR45}              $	&	&	0.708 	&	&	$<$1E-16	&	&	30.761 	\\
5	&	&	$	{\rm PLEC\_Flux\_Density}$	&	&	0.688 	&	&	$<$1E-16	&	&	29.138 	\\
6	&	&	$	{\rm HR34}              $	&	&	0.658 	&	&	$<$1E-16	&	&	23.982 	\\
7	&	&	$	{\rm HR56	}         	$	&	&	0.657 	&	&	$<$1E-16	&	&	21.482 	\\
8	&	&	$	{\rm nuFnu\_Band7}		$	&	&	0.600 	&	&	$<$1E-16	&	&	15.884 	\\
9	&	&	$	{\rm Flux\_Band2}		$	&	&	0.562 	&	&	$<$1E-16	&	&	7.420 	\\
10	&	&	$	{\rm Flux\_Band7}		$	&	&	0.564 	&	&	$<$1E-16	&	&	12.496 	\\
11	&	&	$	{\rm nuFnu\_Band2}		$	&	&	0.554 	&	&	$<$1E-16	&	&	7.354 	\\
12	&	&	$	{\rm PLEC\_Expfactor}	$	&	&	0.545 	&	&	$<$1E-16	&	&	8.531 	\\
13	&	&	$	{\rm Frac\_Variability} $	&	&	0.530 	&	&	$<$1E-16	&	&	9.742 	\\
14	&	&	$	{\rm HR67}  			$	&	&	0.504 	&	&	$<$1E-16	&	&	3.819 	\\
15	&	&	$	{\rm Variability\_Index}$	&	&	0.478 	&	&	$<$1E-16	&	&	6.189 	\\
16	&	&	$	{\rm Flux\_Band3}		$	&	&	0.467 	&	&	$<$1E-16	&	&	4.192 	\\
17	&	&	$	{\rm Npred} 			$	&	&	0.460 	&	&	$<$1E-16	&	&	4.101 	\\
18	&	&	$	{\rm nuFnu\_Band3}		$	&	&	0.447 	&	&	$<$1E-16	&	&	3.190 	\\
19	&	&	$	{\rm nuFnu\_Band6}		$	&	&	0.427 	&	&	$<$1E-16	&	&	8.957 	\\
20	&	&	$	{\rm Flux\_Band6}		$	&	&	0.409 	&	&	$<$1E-16	&	&	5.787 	\\
21	&	&	$	{\rm HR23}	    		$	&	&	0.378 	&	&	$<$1E-16	&	&	2.136 	\\
22	&	&	$	{\rm Flux\_Band1}		$	&	&	0.342 	&	&	$<$1E-16	&	&	2.294 	\\
23	&	&	$	{\rm nuFnu\_Band1}		$	&	&	0.341 	&	&	$<$1E-16	&	&	2.154 	\\
 \hline 
24	&	&	$	{\rm HS123} 			$	&	&	0.244 	&	&	$<$1E-16	&	&	2.165 	\\
25	&	&	$	{\rm HS234} 			$	&	&	0.247 	&	&	$<$1E-16	&	&	2.528 	\\
26	&	&	$	{\rm LP\_beta}   		$	&	&	0.256 	&	&	$<$1E-16	&	&	2.918 	\\
27	&	&	$	{\rm Energy\_Flux100}	$	&	&	0.233 	&	&	$<$1E-16	&	&	2.378 	\\
28	&	&	$	{\rm Flux\_Band4}		$	&	&	0.227 	&	&	3.33E-16	&	&	1.921 	\\
29	&	&	$	{\rm LP\_SigCurv}		$	&	&	0.237 	&	&	$<$1E-16	&	&	2.790 	\\
30	&	&	$	{\rm nuFnu\_Band4}		$	&	&	0.212 	&	&	3.02E-14	&	&	2.052 	\\
31	&	&	$	{\rm HR12}  			$	&	&	0.211 	&	&	4.46E-14	&	&	2.042 	\\
32	&	&	$	{\rm PLEC\_SigCurv} 	$	&	&	0.208 	&	&	1.01E-13	&	&	2.839 	\\
33	&	&	$	{\rm HS345} 			$	&	&	0.179 	&	&	2.46E-10	&	&	1.942 	\\
34	&	&	$	{\rm HS456} 			$	&	&	0.165 	&	&	8.89E-09	&	&	2.466 	\\
35	&	&	$	{\rm Flux1000}   		$	&	&	0.142 	&	&	1.23E-06	&	&	2.384 	\\
36	&	&	$	{\rm nuFnu\_Band5}		$	&	&	0.142 	&	&	1.28E-06	&	&	3.128 	\\
37	&	&	$	{\rm Signif\_Avg}		$	&	&	0.121 	&	&	5.77E-05	&	&	2.992 	\\
38	&	&	$	{\rm Flux\_Band5}		$	&	&	0.120 	&	&	6.95E-05	&	&	2.362 	\\
39	&	&	$	{\rm nuFnu\_syn}		$	&	&	0.121 	&	&	6.33E-05	&	&	3.616 	\\
40	&	&	$	{\rm HS567} 			$	&	&	0.107 	&	&	5.62E-04	&	&	2.089 	\\
 \hline 
41	&	&	$	{\rm PLEC\_Exp\_Index}	$	&	&	0.001 	&	&	1.00E+00	&	&	0.000 	\\
42	&	&	$	{\rm LP\_Index}	        $	&	&	0.793 	&	&	$<$1E-16	&	&	52.481 	\\
43	&	&	$	{\rm LP\_Flux\_Density}	$	&	&	0.685 	&	&	$<$1E-16	&	&	21.968 	\\
44	&	&	$	{\rm PL\_Flux\_Density}	$	&	&	0.684 	&	&	$<$1E-16	&	&	24.419 	\\
45	&	&	$	{\rm PLEC\_Index}		$	&	&	0.586 	&	&	$<$1E-16	&	&	10.554 	\\
 \hline 
\end{tabular}
\label{Tab1}\\
Note: Column 1 presents the parameter labels in the sample. Column 2 lists the selected parameters. The two-sample Kolmogorov-Smirnov test results for the test statistic (D) and the p-value($p$) are presented in Columns 3 and 4, respectively. The mean decrease  Gini coefficient (Gini), an indicator of variable importance in RFs are presented in Column 5.
\end{table*}

The Fourth Catalog of Active Galactic Nuclei Detected by the Fermi Large Area Telescope  \citep{2020ApJ...892..105A} -- Data Release 2 (4LAC-DR2\footnote{\url{https://fermi.gsfc.nasa.gov/ssc/data/access/lat/4LACDR2/}} was released on  October 16,  2020, \citealt{2020arXiv201008406L}) includes 3131 sources, with 3063 blazars (707 FSRQs, 1236 BL Lacs, 1120 blazar candidates of unknown types, BCUs), and 68 other AGNs,  located at high Galactic latitudes ($|b|>10\deg$), of which 3063 blazars include 1388 LSP, 474 ISP, 506 HSP, and 695 no SED class. From the high Galactic latitudes 4LAC-DR2 FITS tables: “table-4LAC-DR2-h.fits”\footnote{\url{https://fermi.gsfc.nasa.gov/ssc/data/access/lat/4LACDR2/table-4LAC-DR2-h.fits}}, we select 1680 blazars that have known optical classifications (FSRQs and BL Lacs) and SED-based classifications (LSP, ISP, and HSP), which include 651 FSRQs and 1029 BL Lacs (960 LSP, 334 ISP and 386 HSP). The 1680 blazars are divided into three subsamples: training, validation, and forecast. Where, the 651 FSRQs and 701 (ISP and HSP) BL Lacs are viewed as the training and validation samples, while 328 LSP BL Lacs are viewed as a forecast sample.

 \vspace{2mm}

We selected the data from 
the 4LAC-DR2 catalog \citep{2020arXiv201008406L} and  
the 4FGL-DR2 catalog \citep{2020arXiv200511208B}. The 4LAC-DR2 FITS table (“table-4LAC-DR2-h.fits”) lists 37 variables (37 columns). In the 4FGL-DR2 FITS table (“gll\_psc\_v27.fit”) of the 4FGL-DR2 catalog, 74 variables are reported using 142 columns (also see Table 12 of  \citealt{2020ApJS..247...33A}). 
Among the 74  variables, some variables contain multiple columns. 
For instance, the parameters: 
“Flux\_Band”
with seven columns are used to present the integral photon flux in each of the seven spectral bands that are labeled as Flux\_Band1, Flux\_Band2  ... etc.
“nuFnu{\_Band}”
with seven columns are used to present the SED for the spectral bands, labeled as nuFnu{\_Band}1, nuFnu{\_Band}2  ... etc. (see Table \ref{Tab1});
and so on.
For the description of other multi-column parameters can be referenced in  \cite{2020ApJS..247...33A} or \cite{2019ApJ...887..134K}.

\vspace{2mm}

In addition to all the parameters (data columns) reported in 4FGL-DR2 and 4LAC-DR2, following \cite{2012ApJ...753...83A}, 
similar to \cite{2014ApJ...782...41D}, \cite{2016ApJ...820....8S}, and \cite{2021ApJ...916...93Z}, 
the hardness ratios are calculated using the following Equation:
\begin{equation}\label{equ1}
HR_{ij} = \frac{\nu{F}\nu_j - \nu{F}\nu_i}{\nu{F}\nu_j + \nu{F}\nu_i}
\end{equation}
where $i$ and $j$ are indices corresponding to the seven different spectral energy bands defined in the 4FGL-DR2 catalog: 
$i,j=1$: 50 $-$ 100MeV;
2: 100 $-$ 300MeV;
3: 300MeV $-$ 1GeV;
4: 1 $-$ 3GeV;
5: 3 $-$ 10GeV;
6: 10 $-$ 30GeV;
and 7: 30 $-$ 300 GeV.
Combining two hardness ratios, a hardness curvature parameter was also constructed as:
\begin{equation}\label{equ2}
HRC_{ijk} = HR_{ij} - HR_{jk}
\end{equation}
for instance, HRC$_{234}$ = HR$_{23}$-HR$_{34}$, where 2, 3, and 4 are indices for 2: 100 $-$ 300MeV; 3: 300MeV $-$ 1GeV; 4: 1 $-$ 3GeV; respectively.

\vspace{2mm}

Based on the VLBI counterpart listed in the 4LAC-DR2 catalog \citep[see][]{2020arXiv201008406L}, 
and using the TOPCAT\footnote{\url{http://www.star.bris.ac.uk/~mbt/topcat/}} software \citep{2005ASPC..347...29T}, 
we cross-matched the latest version ("rfc\_2022a", as of May 20, 2022) of the Radio Fundamental Catalog (RFC)\footnote{\url{http://astrogeo.org/sol/rfc/rfc_2022a/}}, 
 which is the most complete catalogue of positions of compact radio sources and lists 20,499 sources  
(see \citealt{2002ApJS..141...13B,2003AJ....126.2562F,2021AJ....161...14P} and references therein),
 to obtain the flux density (in units of Jy) of the S-band, C-band, X-band ($F_{R}$), U-band, and K-band. 
 Among the 1680 selected sources, 
there are 877, 785, 1483, 374, and 337 radio data points in S-band, C-band, X-band, U-band, and K-band respectively.
 The number of matches in the X-band is the largest, including 1185 (1185/1352 $\simeq$ 87.66\%) sources with observational data (and still 167 source data missing) in the training and validation sample, and 298 (298/328 $\simeq$ 90.85\%) sources with observational data (and still 30 source data missing) in the forecast sample.
Where, in this work,
for the missing data in the random forest calculation,
it is filled with the median using the $na.roughfix()$ function of the random forests algorithm.

\vspace{2mm}

Similar to the parameter selection rules  of previous work (see \citealt{2019ApJ...887..134K} for the detailed description), 
firstly, the subset of parameters and their associated data are identified. Where, the coordinate columns, error columns, string columns, and most data missing columns are removed; Keeping one of the same data columns from the 4LAC table and the 4FGL table; Merging the defined data (e.g., ``$HR_{ij}$, $HRC_{ijk}$”) that are created above using equation \ref{equ1} and \ref{equ2}; And for the VLBI radio data, only the X-band flux density was chosen because there are too many missing data for other bands. The 45 candidate parameters were preliminarily selected (see Table \ref{Tab1})  from the 4LAC table,  4FGL table,  created data and  RFC  data.

\vspace{2mm}

Secondly, in order to simplify the calculation, some parameters are pre-selected for the SML algorithms. The Two Sample Kolmogorov-Smirnov test (e.g., \citealt{2018MNRAS.475.1708A,2019ApJ...872..189K,2019ApJ...887..134K,2020ApJ...891...87K}) is applied to two subsamples of the data (651 FSRQs and  701 ISP (and HSP) BL Lacs) to calculate the independence of the 45 parameters, the results are summarized in Table \ref{Tab1}. The Mean Decrease Gini coefficients (Gini coefficients) that is a measure of how each variable contributes to the homogeneity of the nodes and leaves in the resulting random forest \citep[e.g.][]{MeanDecreaseGini,10.1371/journal.pone.0230799}. The higher the value of mean decrease Gini score, the higher the importance of the variable in the model. Which is an established method to determine the variables importance that is defined for each variable as the sum of the decrease in impurity for each of the surrogate variables at each node in the book of classification and regression trees (see \citealt{Breiman1984ClassificationAR,R_randomForest} for the details and references therein). The Mean Decrease Gini coefficients  are also computed  by applying a random forests algorithm (see Section  \ref{sec:method}) to the  45 parameters' data. The results are consistent with those of the two sample K-S tests and  are also presented in Table \ref{Tab1}. Considering p $>$ 0.05\footnote{where, p $>$ 0.05 indicates that the two populations should be the same distribution, which does not reject the null hypothesis} and  the Gini coefficients (Gini $\simeq$ 0.000), one parameter (``PLEC\_Exp\_Index”) is excluded; Comparing D-values in K-S test and  the Gini coefficients in random forests, for the similar or identical parameters (see Table \ref{Tab1}), the four parameters with smaller D-values and Gini coefficients: LP\_Index and PLEC\_Index, or PL\_Flux\_Density, LP\_Flux\_Density, are also excluded;  PL\_Index and  PLEC\_Flux\_Density with a bigger D-values and Gini coefficients are selected. Therefore,  40 parameters are selected in this work.

\vspace{2mm}

For the selected 40 parameters, there are 1.099512E$+$12 different combinations, which need to costs too long time to utilize the random forests to calculate each combination of parameters. This is not possible for us to accomplish. In order to reduce the calculation time, to ensure the study can be completed (see \citealt{2019ApJ...887..134K} for the detailed description), finally, we further sub-selected 23 parameters by considering D $>$ {0.3} in the K-S test and Gini $>$ 2.1 in random forests algorithm. A simple horizontal line is introduced in the table to distinguish the collection of 23 parameters utilized. Based on the  selected  23 parameters, a subset of the data sample is selected from the 4FGL table, 4LAC table, created data and RFC data, which includes {1680} blazars (651  FSRQs, 701 ISP and HSP BL Lacs, and 328 LSP BL Lacs). These 328 LSP BL Lacs are listed in Table \ref{Tab3}.

\vspace{3mm}
\section{Optimal combinations of parameters and Results} \label{Result_opt}
\vspace{3mm}

In the random forest calculation, for the training and  validation sample, 
approximately 4/5 of 1352 blazars (651 FSRQs and 701 HSP and ISP BL Lacs) are randomly (random seed = 123) assigned to the training sample, and the remaining ones (e.g., approximately 1/5) are considered as the validation sample. Here, the training sample include 1082 blazars (528 FSRQs and  554 ISP, or HSP BL Lacs), and the validation sample has  270 blazars (123 FSRQs and  147 ISP, or HSP BL Lacs).

\begin{figure*} 
\centering
\includegraphics[width=14cm,height=11cm]{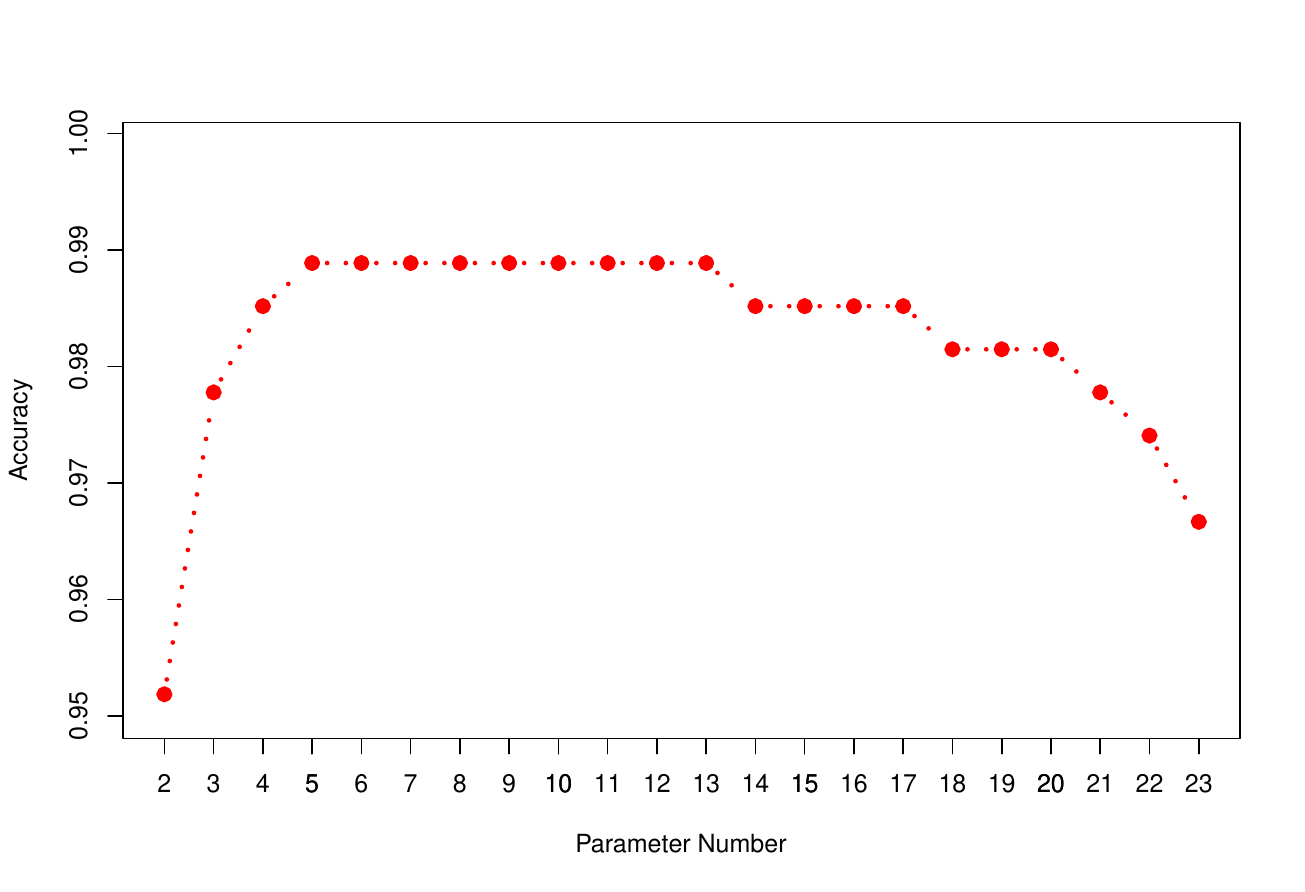}
\caption{Highest accuracy for the different number of combinations of parameters in Random Forests algorithm. \label{FigA1}}
\end{figure*}

\begin{figure*} 
\centering
\includegraphics[width=17cm,height=8cm]{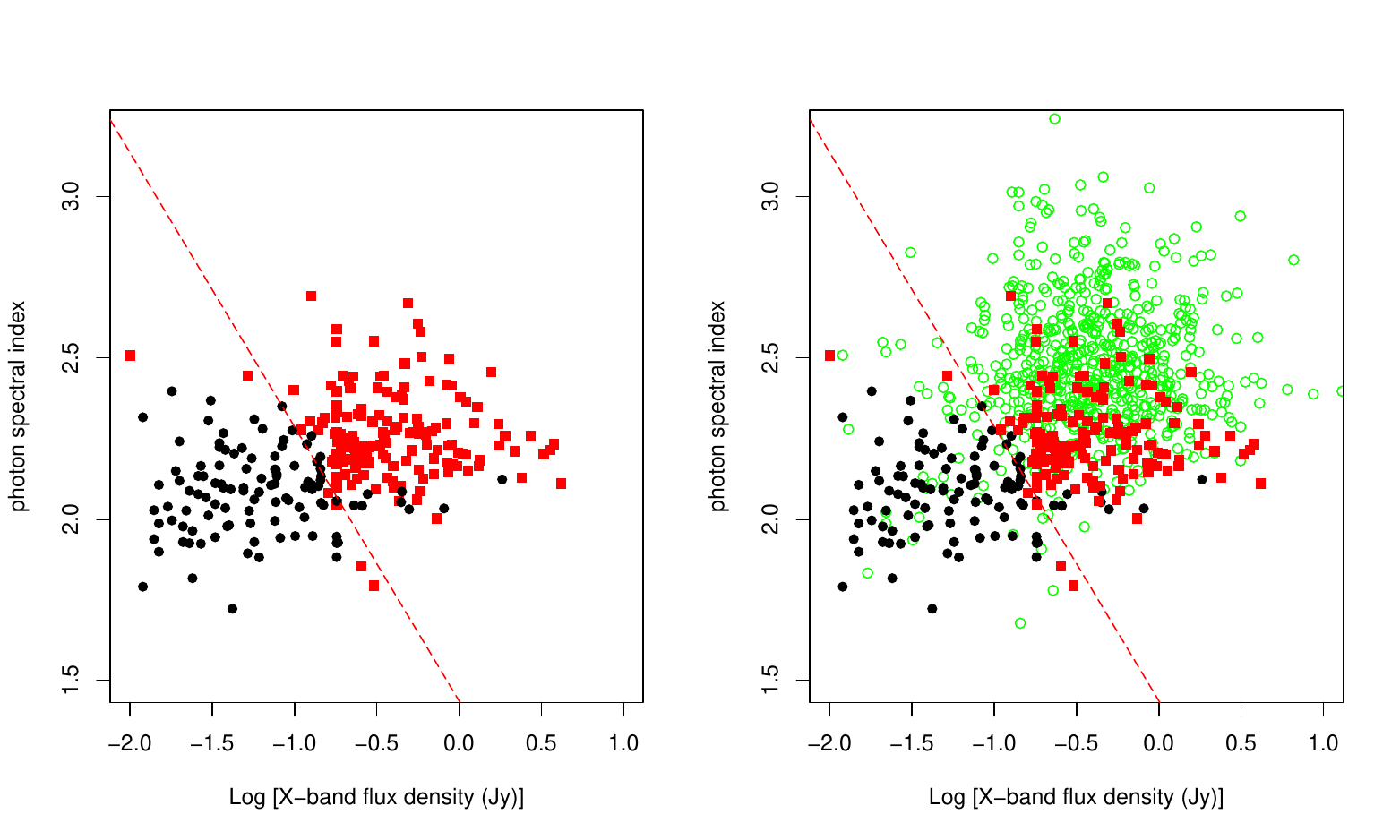}
\caption{Classification scatterplots for the Fermi $\gamma$-ray photon spectral index ($\Gamma_{\rm ph}$) and  the X-band VLBI radio flux (log$F_{R}$), where the black filled circles, red solid squares, and  green empty circles indicate TBLs, FBLs and FSRQs respectively. \label{Fig_line}}
\end{figure*}



\begin{table*}
	\centering
	\caption{The test accuracy, predict results , and parameters for the optimal combinations in RF algorithm.}
	\label{tab_resultA}
	\begin{tabular}{lcccccccccccccccc} 
		\hline\hline
{N}& {$N_{\rm TBLLs}$} & {$N_{\rm FBLLs}$} &{Acc} &{p1}& {p2}& {p3}& {p4} &{p5} & {p6} &{p7} & {p8} &{p9} & {p10} & {p11} & {p12} & {p13} \\
{(1)}& {(2)} & {(3)} &{(4)} &{(5)}& {(6)}& {(7)}& {(8)} &{(9)} & {(10)} &{(11)} & {(12)} &{(13)} & {(14)} & {(15)} & {(16)}  & {(17)} \\
		\hline

5$^*$&	137	&	191	&	0.9889 	&	2	&	4	&	5	&	7	&	13	&	...	&	...	&	...	&	...	&	...	&	...	&	...	&	...	\\
\cline{5-9} 
6$^*$&	138	&	190	&	0.9889 	&	2	&	4	&	7	&	8	&	18	&	19	&	...	&	...	&	...	&	...	&	...	&	...	&	...	\\
\cline{5-10} 
6      &	140	&	188	&	0.9889 	&	2	&	4	&	5	&	7	&	13	&	18	&	...	&	...	&	...	&	...	&	...	&	...	&	...	\\
6	&	141	&	187	&	0.9889 	&	2	&	4	&	5	&	7	&	13	&	20	&	...	&	...	&	...	&	...	&	...	&	...	&	...	\\
6	&	139	&	189	&	0.9889 	&	2	&	4	&	5	&	12	&	13	&	19	&	...	&	...	&	...	&	...	&	...	&	...	&	...	\\
6	&	140	&	188	&	0.9889 	&	2	&	4	&	8	&	13	&	16	&	19	&	...	&	...	&	...	&	...	&	...	&	...	&	...	\\
7$^*$&	140	&	188	&	0.9889 	&	2	&	4	&	5	&	7	&	9	&	13	&	18	&	...	&	...	&	...	&	...	&	...	&	...	\\
\cline{5-11} 
7	&	141	&	187	&	0.9889 	&	2	&	4	&	5	&	7	&	12	&	13	&	15	&	...	&	...	&	...	&	...	&	...	&	...	\\
7	&	142	&	186	&	0.9889 	&	2	&	4	&	5	&	7	&	12	&	13	&	19	&	...	&	...	&	...	&	...	&	...	&	...	\\
7	&	143	&	185	&	0.9889 	&	2	&	4	&	5	&	7	&	12	&	13	&	20	&	...	&	...	&	...	&	...	&	...	&	...	\\
7	&	137	&	191	&	0.9889 	&	2	&	4	&	5	&	7	&	13	&	15	&	17	&	...	&	...	&	...	&	...	&	...	&	...	\\
7	&	138	&	190	&	0.9889 	&	2	&	4	&	5	&	7	&	13	&	15	&	18	&	...	&	...	&	...	&	...	&	...	&	...	\\
7	&	139	&	189	&	0.9889 	&	2	&	4	&	5	&	7	&	13	&	15	&	19	&	...	&	...	&	...	&	...	&	...	&	...	\\
7	&	142	&	186	&	0.9889 	&	2	&	4	&	5	&	7	&	13	&	15	&	20	&	...	&	...	&	...	&	...	&	...	&	...	\\
7	&	141	&	187	&	0.9889 	&	2	&	4	&	5	&	7	&	13	&	16	&	18	&	...	&	...	&	...	&	...	&	...	&	...	\\
7	&	143	&	185	&	0.9889 	&	2	&	4	&	5	&	7	&	13	&	16	&	19	&	...	&	...	&	...	&	...	&	...	&	...	\\
7	&	139	&	189	&	0.9889 	&	2	&	4	&	5	&	7	&	13	&	20	&	22	&	...	&	...	&	...	&	...	&	...	&	...	\\
7	&	128	&	200	&	0.9889 	&	2	&	8	&	10	&	13	&	14	&	18	&	19	&	...	&	...	&	...	&	...	&	...	&	...	\\
7	&	141	&	187	&	0.9889 	&	2	&	4	&	8	&	13	&	14	&	19	&	21	&	...	&	...	&	...	&	...	&	...	&	...	\\
7	&	138	&	190	&	0.9889 	&	2	&	4	&	8	&	13	&	14	&	20	&	23	&	...	&	...	&	...	&	...	&	...	&	...	\\
8$^*$&	146	&	182	&	0.9889 	&	2	&	4	&	7	&	8	&	13	&	18	&	19	&	20	&	...	&	...	&	...	&	...	&	...	\\
\cline{5-12} 
8	&	142	&	186	&	0.9889 	&	2	&	4	&	5	&	7	&	8	&	13	&	14	&	19	&	...	&	...	&	...	&	...	&	...	\\
8	&	140	&	188	&	0.9889 	&	2	&	4	&	5	&	7	&	9	&	12	&	13	&	19	&	...	&	...	&	...	&	...	&	...	\\
8	&	137	&	191	&	0.9889 	&	2	&	4	&	5	&	7	&	9	&	12	&	13	&	20	&	...	&	...	&	...	&	...	&	...	\\
8	&	141	&	187	&	0.9889 	&	2	&	4	&	5	&	7	&	9	&	13	&	18	&	23	&	...	&	...	&	...	&	...	&	...	\\
8	&	141	&	187	&	0.9889 	&	2	&	4	&	5	&	7	&	11	&	12	&	13	&	20	&	...	&	...	&	...	&	...	&	...	\\
8	&	140	&	188	&	0.9889 	&	2	&	4	&	5	&	7	&	11	&	13	&	16	&	22	&	...	&	...	&	...	&	...	&	...	\\
8	&	143	&	185	&	0.9889 	&	2	&	4	&	5	&	7	&	12	&	13	&	14	&	19	&	...	&	...	&	...	&	...	&	...	\\
9$^*$&	136	&	192	&	0.9889 	&	2	&	4	&	6	&	8	&	12	&	13	&	14	&	19	&	20	&	...	&	...	&	...	&	...	\\
\cline{5-13} 
9	&	136	&	192	&	0.9889 	&	2	&	4	&	6	&	8	&	12	&	13	&	18	&	19	&	20	&	...	&	...	&	...	&	...	\\
9	&	136	&	192	&	0.9889 	&	2	&	4	&	7	&	10	&	13	&	14	&	18	&	19	&	20	&	...	&	...	&	...	&	...	\\
9	&	134	&	194	&	0.9889 	&	2	&	4	&	10	&	13	&	14	&	19	&	20	&	22	&	23	&	...	&	...	&	...	&	...	\\
9	&	135	&	193	&	0.9889 	&	2	&	4	&	5	&	7	&	8	&	10	&	13	&	20	&	21	&	...	&	...	&	...	&	...	\\
9	&	133	&	195	&	0.9889 	&	2	&	4	&	8	&	10	&	12	&	13	&	14	&	18	&	19	&	...	&	...	&	...	&	...	\\
10$^*$&	139	&	189	&	0.9889 	&	2	&	4	&	5	&	7	&	8	&	12	&	13	&	15	&	19	&	22	&	...	&	...	&	...	\\
\cline{5-14} 
11$^*$&	136	&	192	&	0.9889 	&	2	&	4	&	7	&	8	&	9	&	10	&	13	&	14	&	19	&	20	&	22	&	...	&	...	\\
\cline{5-15} 
12$^*$&	143	&	185	&	0.9889 	&	2	&	4	&	5	&	7	&	12	&	13	&	15	&	18	&	19	&	20	&	21	&	22	&	...	\\
\cline{5-16} 
13$^*$&	144	&	184	&	0.9889 	&	2	&	4	&	5	&	7	&	8	&	12	&	13	&	14	&	15	&	19	&	20	&	21	&	22	\\
\cline{5-17} 
...	&	...	&	...	&	...      	&	...	&	...	&	...	&	...	&	...	&	...	&	...	&	...	&	...	&	...	&	...	&	...	&	...	\\
\hline
	\end{tabular}\\
Note: The number of parameters for the optimal combination are presented in Column 1. 
The highest accuracies of each classifier are presented in Column 4. 
The number of true BL Lacs (TBLLs) and false BL Lacs (FBLLs) that possible intrinsically FSRQs predicted by Random Forests algorithm with the default settings for the LSP BL Lacs (predicted dataset) are presented in Columns 2 and 3. 
The labels of the parameters are presented in Columns 5-17, these correspond to the labels in Table \ref{Tab1}, Column 1. 
Here, the one combination for the  different  number of parameters (e.g., $10^*$, $11^*$, $12^*$, and $13^*$) for the optimal combinations is shown for guidance regarding its form and content.  (This table is available in its entirety in machine-readable form.)
\end{table*}

\begin{table*}
	\centering
	\caption{The Number of the Optimal Parameters and Combinations.}
	\label{tab_numb}
	\begin{tabular}{cccccccccc} 
\hline\hline
{Algorithm} &{5Pars} & {6Pars} &{7Pars} &  {8Pars} & {9Pars} &{10Pars} &{11Pars}  &{12Pars}  &{13Pars} \\
\hline
RFs  &   1  &   5  &   14  &   35  &   52  &   39  &   28  &   2  &    2 	           \\
 \hline
	\end{tabular}\\
Note: The Algorithm  are presented in Column 1. 
The number of combinations with the highest accuracies in RFs algorithm for 5$-$13 parameters are presented in Columns 2$-$10 
(see the machine-readable format of Table \ref{tab_resultA} for the details).
\end{table*}


\begin{table*}
	\centering
	\tiny
	\caption{The predicted classification results of Fermi LSP BL Lacs.}
	\label{Tab3}
	\begin{tabular}{cccccccccccrcccc} 
\hline\hline
{4FGL name}  & {RF5} & {RF6} & {RF7} & {RF8} & {RF9} & {RF10} & {RF11} & {RF12} & {RF13} & {$C_{9}$} & {$M_{Fan}$}  & {$M_{CKZ}$} & {CD} & {$M_{\rm wise}$} \\
{(1)}& {(2)} & {(3)} &{(4)} &{(5)}& {(6)}& {(7)}& {(8)}           &{(9)}           & {(10)}           &{(11)}      & {(12)}      &{(13)}      & {(14)}      & {(15)}  \\
\hline
4FGL J0001.2$-$0747	&	fsrq	&	fsrq	&	fsrq	&	bll	&	bll	&	fsrq	&	fsrq	&	fsrq	&	bll	&	UNK	&	...	&	...	&	...	&	...	\\
4FGL J0003.2$+$2207	&	bll	&	bll	&	bll	&	bll	&	bll	&	bll	&	fsrq	&	bll	&	bll	&	UNK	&	...	&	...	&	...	&	...	\\
4FGL J0003.9$-$1149	&	fsrq	&	fsrq	&	fsrq	&	fsrq	&	fsrq	&	fsrq	&	fsrq	&	fsrq	&	fsrq	&	fsrq	&	...	&	...	&	...	&	...	\\
4FGL J0006.3$-$0620	&	fsrq	&	fsrq	&	fsrq	&	fsrq	&	fsrq	&	fsrq	&	fsrq	&	fsrq	&	fsrq	&	fsrq	&	...	&	...	&	...	&	BZQ	\\
4FGL J0008.0$+$4711	&	bll	&	bll	&	bll	&	bll	&	bll	&	bll	&	bll	&	bll	&	bll	&	bll	&	...	&	...	&	...	&	...	\\
4FGL J0009.1$+$0628	&	bll	&	bll	&	bll	&	bll	&	bll	&	bll	&	bll	&	bll	&	bll	&	bll	&	...	&	...	&	...	&	...	\\
4FGL J0013.1$-$3955	&	fsrq	&	fsrq	&	fsrq	&	fsrq	&	fsrq	&	fsrq	&	fsrq	&	fsrq	&	fsrq	&	fsrq	&	...	&	...	&	...	&	...	\\
4FGL J0014.1$+$1910	&	fsrq	&	fsrq	&	fsrq	&	fsrq	&	fsrq	&	fsrq	&	fsrq	&	fsrq	&	fsrq	&	fsrq	&	...	&	...	&	1.41	&	...	\\
4FGL J0019.6$+$2022	&	fsrq	&	fsrq	&	fsrq	&	fsrq	&	fsrq	&	fsrq	&	fsrq	&	fsrq	&	fsrq	&	fsrq	&	...	&	...	&	...	&	...	\\
4FGL J0022.1$-$1854	&	bll	&	bll	&	bll	&	bll	&	bll	&	bll	&	bll	&	bll	&	bll	&	bll	&	...	&	...	&	...	&	...	\\
4FGL J0022.5$+$0608	&	bll	&	fsrq	&	bll	&	fsrq	&	fsrq	&	fsrq	&	fsrq	&	bll	&	fsrq	&	UNK	&	...	&	...	&	...	&	...	\\
4FGL J0023.9$+$1603	&	bll	&	bll	&	bll	&	bll	&	bll	&	bll	&	bll	&	bll	&	bll	&	bll	&	...	&	...	&	...	&	...	\\
4FGL J0029.0$-$7044	&	fsrq	&	fsrq	&	fsrq	&	bll	&	bll	&	fsrq	&	fsrq	&	bll	&	fsrq	&	UNK	&	...	&	...	&	...	&	...	\\
4FGL J0032.4$-$2849	&	bll	&	fsrq	&	bll	&	fsrq	&	fsrq	&	fsrq	&	bll	&	fsrq	&	fsrq	&	UNK	&	...	&	...	&	1.26	&	...	\\
4FGL J0035.8$-$0837	&	bll	&	bll	&	bll	&	bll	&	bll	&	bll	&	bll	&	bll	&	bll	&	bll	&	...	&	...	&	...	&	...	\\
4FGL J0038.1$+$0012	&	bll	&	bll	&	bll	&	bll	&	bll	&	bll	&	bll	&	bll	&	bll	&	bll	&	...	&	...	&	...	&	...	\\
4FGL J0040.3$+$4050	&	bll	&	bll	&	bll	&	bll	&	bll	&	bll	&	bll	&	bll	&	bll	&	bll	&	...	&	...	&	...	&	...	\\
4FGL J0049.7$+$0237	&	fsrq	&	fsrq	&	fsrq	&	fsrq	&	fsrq	&	fsrq	&	fsrq	&	fsrq	&	fsrq	&	fsrq	&	Fan\_fsrq	&	...	&	...	&	...	\\
4FGL J0056.8$+$1626	&	fsrq	&	fsrq	&	fsrq	&	fsrq	&	fsrq	&	fsrq	&	fsrq	&	fsrq	&	fsrq	&	fsrq	&	...	&	...	&	...	&	BZQ	\\
4FGL J0058.0$-$3233	&	bll	&	bll	&	bll	&	bll	&	bll	&	bll	&	bll	&	bll	&	bll	&	bll	&	...	&	...	&	...	&	...	\\
4FGL J0105.1$+$3929	&	fsrq	&	fsrq	&	fsrq	&	fsrq	&	fsrq	&	fsrq	&	fsrq	&	fsrq	&	fsrq	&	fsrq	&	...	&	...	&	1.95	&	...	\\
4FGL J0107.4$+$0334	&	fsrq	&	fsrq	&	fsrq	&	fsrq	&	fsrq	&	fsrq	&	fsrq	&	fsrq	&	fsrq	&	fsrq	&	...	&	...	&	...	&	...	\\
4FGL J0112.1$+$2245	&	bll	&	bll	&	bll	&	bll	&	bll	&	bll	&	bll	&	bll	&	bll	&	bll	&	...	&	...	&	...	&	...	\\
4FGL J0113.7$+$0225	&	fsrq	&	fsrq	&	fsrq	&	fsrq	&	fsrq	&	fsrq	&	fsrq	&	fsrq	&	fsrq	&	fsrq	&	...	&	...	&	...	&	...	\\
4FGL J0124.8$-$0625	&	fsrq	&	bll	&	fsrq	&	fsrq	&	fsrq	&	fsrq	&	fsrq	&	fsrq	&	fsrq	&	UNK	&	...	&	...	&	...	&	...	\\
4FGL J0125.3$-$2548	&	fsrq	&	fsrq	&	fsrq	&	fsrq	&	fsrq	&	fsrq	&	fsrq	&	fsrq	&	fsrq	&	fsrq	&	...	&	...	&	...	&	...	\\
4FGL J0127.2$-$0819	&	bll	&	bll	&	bll	&	bll	&	bll	&	bll	&	bll	&	bll	&	bll	&	bll	&	...	&	...	&	...	&	...	\\
4FGL J0141.4$-$0928	&	fsrq	&	fsrq	&	fsrq	&	fsrq	&	fsrq	&	fsrq	&	fsrq	&	fsrq	&	fsrq	&	fsrq	&	Fan\_fsrq	&	...	&	...	&	...	\\
4FGL J0142.7$-$0543	&	bll	&	bll	&	bll	&	fsrq	&	fsrq	&	bll	&	fsrq	&	bll	&	bll	&	UNK	&	...	&	...	&	...	&	...	\\
4FGL J0144.6$+$2705	&	fsrq	&	fsrq	&	fsrq	&	fsrq	&	fsrq	&	fsrq	&	fsrq	&	fsrq	&	fsrq	&	fsrq	&	...	&	...	&	...	&	...	\\
4FGL J0148.6$+$0127	&	bll	&	bll	&	bll	&	bll	&	bll	&	bll	&	bll	&	bll	&	bll	&	bll	&	...	&	...	&	...	&	...	\\
4FGL J0202.7$+$4204	&	fsrq	&	fsrq	&	fsrq	&	fsrq	&	fsrq	&	fsrq	&	fsrq	&	fsrq	&	fsrq	&	fsrq	&	...	&	...	&	...	&	...	\\
4FGL J0203.6$+$7233	&	fsrq	&	fsrq	&	fsrq	&	fsrq	&	fsrq	&	fsrq	&	fsrq	&	fsrq	&	fsrq	&	fsrq	&	...	&	...	&	...	&	...	\\
4FGL J0203.7$+$3042	&	fsrq	&	fsrq	&	fsrq	&	fsrq	&	fsrq	&	fsrq	&	fsrq	&	fsrq	&	fsrq	&	fsrq	&	...	&	...	&	2.34	&	...	\\
4FGL J0208.3$-$6838	&	bll	&	bll	&	bll	&	bll	&	bll	&	bll	&	bll	&	bll	&	bll	&	bll	&	...	&	...	&	...	&	...	\\
4FGL J0208.5$-$0046	&	fsrq	&	fsrq	&	fsrq	&	fsrq	&	fsrq	&	fsrq	&	fsrq	&	fsrq	&	fsrq	&	fsrq	&	...	&	...	&	...	&	BZQ	\\
4FGL J0209.9$+$7229	&	fsrq	&	fsrq	&	fsrq	&	fsrq	&	fsrq	&	fsrq	&	fsrq	&	fsrq	&	fsrq	&	fsrq	&	...	&	CKZ\_fsrq	&	2.24	&	...	\\
4FGL J0217.2$+$0837	&	fsrq	&	fsrq	&	fsrq	&	fsrq	&	fsrq	&	fsrq	&	fsrq	&	fsrq	&	fsrq	&	fsrq	&	...	&	...	&	...	&	...	\\
4FGL J0219.5$+$0724	&	bll	&	bll	&	bll	&	bll	&	bll	&	bll	&	bll	&	bll	&	bll	&	bll	&	...	&	...	&	...	&	...	\\
4FGL J0224.0$-$7941	&	fsrq	&	fsrq	&	fsrq	&	fsrq	&	fsrq	&	fsrq	&	fsrq	&	fsrq	&	bll	&	UNK	&	...	&	...	&	...	&	...	\\
4FGL J0231.2$-$5754	&	bll	&	bll	&	bll	&	bll	&	bll	&	bll	&	bll	&	bll	&	bll	&	bll	&	...	&	...	&	...	&	...	\\
4FGL J0238.6$+$1637	&	fsrq	&	fsrq	&	fsrq	&	fsrq	&	fsrq	&	fsrq	&	fsrq	&	fsrq	&	fsrq	&	fsrq	&	Fan\_fsrq	&	...	&	2.69	&	...	\\
4FGL J0241.0$-$0505	&	bll	&	fsrq	&	bll	&	bll	&	fsrq	&	bll	&	bll	&	bll	&	bll	&	UNK	&	...	&	...	&	...	&	BZQ	\\
4FGL J0243.4$+$7119	&	bll	&	bll	&	bll	&	bll	&	bll	&	bll	&	bll	&	bll	&	bll	&	bll	&	...	&	...	&	...	&	...	\\
4FGL J0245.1$-$0257	&	bll	&	bll	&	bll	&	bll	&	bll	&	bll	&	bll	&	bll	&	bll	&	bll	&	...	&	...	&	...	&	...	\\
4FGL J0255.8$+$0534	&	bll	&	bll	&	bll	&	bll	&	bll	&	bll	&	bll	&	bll	&	bll	&	bll	&	...	&	...	&	...	&	...	\\
4FGL J0258.1$+$2030	&	bll	&	bll	&	bll	&	bll	&	bll	&	bll	&	bll	&	bll	&	bll	&	bll	&	...	&	...	&	...	&	...	\\
4FGL J0301.0$-$1652	&	fsrq	&	fsrq	&	fsrq	&	fsrq	&	fsrq	&	fsrq	&	fsrq	&	fsrq	&	fsrq	&	fsrq	&	...	&	...	&	...	&	...	\\
4FGL J0312.9$+$3614	&	bll	&	bll	&	bll	&	bll	&	bll	&	bll	&	bll	&	bll	&	bll	&	bll	&	...	&	...	&	...	&	...	\\
4FGL J0314.3$-$5103	&	fsrq	&	bll	&	fsrq	&	bll	&	fsrq	&	bll	&	bll	&	fsrq	&	bll	&	UNK	&	...	&	...	&	...	&	...	\\
4FGL J0316.2$+$0905	&	bll	&	bll	&	bll	&	bll	&	bll	&	bll	&	bll	&	bll	&	bll	&	bll	&	...	&	...	&	...	&	...	\\
4FGL J0334.2$-$3725	&	bll	&	bll	&	bll	&	bll	&	bll	&	bll	&	bll	&	bll	&	bll	&	bll	&	...	&	...	&	...	&	...	\\
4FGL J0334.2$-$4008	&	fsrq	&	fsrq	&	fsrq	&	fsrq	&	fsrq	&	fsrq	&	fsrq	&	fsrq	&	fsrq	&	fsrq	&	Fan\_fsrq	&	...	&	1.78	&	...	\\
4FGL J0340.5$-$2118	&	fsrq	&	fsrq	&	fsrq	&	fsrq	&	fsrq	&	fsrq	&	fsrq	&	fsrq	&	fsrq	&	fsrq	&	...	&	...	&	...	&	...	\\
4FGL J0348.6$-$1609	&	fsrq	&	fsrq	&	fsrq	&	fsrq	&	fsrq	&	fsrq	&	fsrq	&	fsrq	&	fsrq	&	fsrq	&	...	&	...	&	...	&	...	\\
4FGL J0354.7$+$8009	&	fsrq	&	fsrq	&	fsrq	&	fsrq	&	fsrq	&	fsrq	&	fsrq	&	fsrq	&	fsrq	&	fsrq	&	...	&	...	&	...	&	...	\\
4FGL J0359.4$-$2616	&	fsrq	&	fsrq	&	fsrq	&	fsrq	&	fsrq	&	fsrq	&	fsrq	&	fsrq	&	fsrq	&	fsrq	&	...	&	CKZ\_fsrq	&	...	&	BZQ	\\
4FGL J0402.0$-$2616	&	bll	&	bll	&	bll	&	bll	&	bll	&	bll	&	bll	&	bll	&	bll	&	bll	&	...	&	...	&	...	&	...	\\
4FGL J0403.5$-$2437	&	fsrq	&	fsrq	&	fsrq	&	fsrq	&	fsrq	&	fsrq	&	fsrq	&	fsrq	&	fsrq	&	fsrq	&	...	&	CKZ\_fsrq	&	...	&	BZQ	\\
4FGL J0407.5$+$0741	&	fsrq	&	fsrq	&	fsrq	&	fsrq	&	fsrq	&	fsrq	&	fsrq	&	fsrq	&	fsrq	&	fsrq	&	Fan\_fsrq	&	CKZ\_fsrq	&	2.69	&	...	\\
4FGL J0422.3$+$1951	&	fsrq	&	fsrq	&	fsrq	&	bll	&	bll	&	bll	&	fsrq	&	bll	&	bll	&	UNK	&	...	&	...	&	...	&	...	\\
4FGL J0424.7$+$0036	&	fsrq	&	fsrq	&	fsrq	&	fsrq	&	fsrq	&	fsrq	&	fsrq	&	fsrq	&	fsrq	&	fsrq	&	...	&	...	&	...	&	...	\\
4FGL J0424.9$-$5331	&	fsrq	&	fsrq	&	fsrq	&	fsrq	&	fsrq	&	fsrq	&	fsrq	&	fsrq	&	fsrq	&	fsrq	&	...	&	...	&	...	&	...	\\
4FGL J0428.6$-$3756	&	fsrq	&	bll	&	fsrq	&	bll	&	fsrq	&	fsrq	&	fsrq	&	fsrq	&	fsrq	&	UNK	&	Fan\_fsrq	&	...	&	6.31	&	BZQ	\\
4FGL J0433.6$+$2905	&	bll	&	bll	&	bll	&	bll	&	bll	&	bll	&	bll	&	bll	&	bll	&	bll	&	Fan\_fsrq	&	...	&	12.59	&	...	\\
4FGL J0438.9$-$4521	&	fsrq	&	fsrq	&	fsrq	&	fsrq	&	fsrq	&	fsrq	&	fsrq	&	fsrq	&	fsrq	&	fsrq	&	Fan\_fsrq	&	CKZ\_fsrq	&	2.51	&	BZQ	\\
	... 		         	&		... 				&	...	&	...	&	...	&	...	&	...	&	...	&	...	&	...	&	...	&	...		&	...	&	...	&	...	\\
 \hline
	\end{tabular}\\
Note: The 4FGL names are listed in Column 1. 
The predicted classification results using RF algorithm for the different parameter combinations in the work are shown in columns 2-10.
The combined classification results  ($C_9$ predictions) is presented in Column 11.
Column 12 and 13  list the predicted classification results ($M_{Fan}$) in \cite{2019ApJ...879..107F} and ($M_{CKZ}$) in \cite{2022MNRAS.515.2215C}, respectively.
The CD values reported in \cite{2021ApJS..253...46P} are listed in Column 14. 
The  BZQs of WIBRaLS2 catalog reported in \cite{2019ApJS..242....4D} are listed in Column 15.
Table \ref{Tab3} is published in its entirety in the machine-readable format. 
A portion is shown here for guidance regarding its form and content. 
\end{table*}

\vspace{2mm}

For the finally sub-selected 23 parameters of 1680 sources, there are 8388607 different combinations. Then, the optimal parameters combinations (OPCs) are searched based on the training, validation and forecast samples using  the random forests algorithms. The default settings for the random forests classification functions ($randomForest()$ in R code) are used to simplify the calculations. After the predictive models are generated and assessed; an effective predictive model is used to forecast whether a LSP BL Lac belongs to the possible intrinsically FRSQ or the actually BL Lac class based on its predictor variables. The main steps to accomplish this in the R platform are publicly available\footnote{\url{https://github.com/ksj7924/Kang2023mnRcode}}.

\vspace{2mm}

The prediction accuracies of the different parameter combinations in the random forests SML algorithms are computed. The highest prediction accuracies for different combinations of parameters in the random forests algorithms (represented with a red dotted line) are illustrated in Figure \ref{FigA1}. As the number of parameters increases, the accuracy gradually reaches its maximum. Here, with 5,  6,  7,  8,  9, 10, 11, 12, or 13 parameters combinations (see Table \ref{tab_resultA}), the accuracy of the  random forests algorithm reaches its maximum. Where, 178 OPCs in total 8388607 different combinations are hunted. There is 1,  5, 14, 35, 52, 39, 28,  2, or 2 combinations  of 5,  6,  7,  8,  9, 10, 11, 12, or 13 parameters  achieving a maximum accuracy (accuracy $\simeq$ {0.9889}) respectively (see Table \ref{tab_resultA} and \ref{tab_numb}). When more parameters are applied, the accuracy begins to decline, which are consistent with the conclusions of our previous work (\citealt{2019ApJ...887..134K}).

\vspace{2mm}

From the 178 OPCs (see Table \ref{tab_resultA} and \ref{tab_numb}),  we select  nine combinations, one combination in one of the combinations with 5, 6, 7, 8, 9, 10, 11, 12, or 13 parameters, respectively (see Table \ref{tab_resultA} for the parameter with underline and marked $^*$ in the number of parameters, e.g.,  5$^*$). Combined the classification results from the nine OPCs ($C_9$ predictions), 
113 true BL Lacs (TBLs) 
and 157 false BL Lacs  (FBLs) that possible intrinsically FSRQs misclassified as BL Lacs, 
are predicted, where 58 remain without a clear prediction,
for 328 LSP BL Lacs reported in the high Galactic latitudes ($|b|>10\deg$) 4LAC-DR2 catalog. The predicted results of 328 LSP BL Lacs are listed in Table \ref{Tab3}.

 \vspace{2mm}

In the $\Gamma_{\rm ph}$-${\rm log}{F_{R}}$ (Fermi $\gamma$-ray photon spectral index and  the X-band VLBI radio flux) plane for the LSP BL Lacs,  we note that the prediction results, between the 113 TBLs (black dots) and 157 FBLs (red squares), show a clear separation (see Figure \ref{Fig_line}). In the two-dimensional parameters space, a simply phenomenological critical line (e.g., $A{\times}x+B{\times}y+C=0$) can be  employed to roughly separate these two subclasses (TBLs and FBLs) (e.g., see \citealt{2018ApJS..235...39C,2022PASJ...74..239X}). Which (this criterion/line) can be obtained from the Support Vector Machines (SVM, the function $svm()$ of the e1071 package in R, see \citealt{R_e1071} for details) with kernel = ``linear” (other settings with default) in two-dimensional parameters space. The optimal critical lines (e.g., see equation \ref{equ_line} identified as  the dotted-dashed red lines in Figure \ref{Fig_line})  with the accuracy value: 92.86\% are obtained in the $\Gamma_{\rm ph}$-${\rm log}{F_{R}}$ plane:
\begin{equation}\label{equ_line}
3.85 \times \Gamma_{\rm ph} + 3.27 \times {\rm log}{F_{R}} - 5.54 =0.
\end{equation}
Of these, 96 of the 113 {TBLs}  (96/113$\simeq$84.96\%) are in the lower left of the line; 151 of the 157 {FBLs}  (151/157$\simeq$96.18\%)  are in the upper right of the line.

 \vspace{2mm}

\begin{table}
\caption{The median  and  mean of  ${\rm log}{F_{R}}$  and $\Gamma_{\rm ph}$  for TBLs,  FBLs, FSRQs }
\tiny
\centering
\begin{tabular}{ccccccccc}
\hline \hline 
                                 &    \multicolumn{2}{c}{{TBLs}} && \multicolumn{2}{c}{{FBLs}} && \multicolumn{2}{c}{{FSRQ}}  \\ 
                                         \cline{2-3}                                   \cline{5-6}                                    \cline{8-9} 
 Paras             &      mean   &median                && mean   &median                  &&  mean   &median    \\ 
(1)                             &        (2)     &   (3)                     &&    (4)     &   (5)                      & &    (6)     &   (7)       \\ 
\hline 
 ${\rm log}{F_{R}}$ 	&   -1.259   &-1.232	                &&   -0.434 & -0.381	                &&   -0.414&   -0.404	\\
$\Gamma_{\rm ph}$	&     2.089  &2.092	                &&	2.265 & 2.244	                &&     2.472&	2.450	\\
 \hline 
\end{tabular}
\label{Tab_median}\\
Note: Column 1 presents the parameters. 
Column 2 and 3  list the  mean  and median  of TBLs;
Column 4 and 5  list the  mean  and median  of FBLs;
Column 6 and 7  list the  mean  and median  of FSRQs, respectively.
\end{table}

\begin{table*}
	\centering
	\caption{Comparison of the other work's Predictions.}
	\label{tab4_comp}
	\begin{tabular}{cccccccc} 
\hline\hline
{Algorithm} &{Class} & {RF Predictions } &{Fan Predictions} &  
{CKZ Predictions} & {CKZ Predictions} &{Paliya  Predictions} &WIBRaLS2\\
 \hline 
 &N               &  328      &   33 FSRQs   & 47 FSRQs   &      39  FSRQs    &626 CD $>$ 1   &  5089 BZQs   \\ 
\cline{2-8} 
 &M 	          &  328      &   29 	     & 39         &       31          &33 	          &   37        \\
\cline{2-8} 
 &FSRQ	 	 	  &  157      &   24	     & 38         &       30          &22 	          &   33        \\
RF&BLL   	 	  &  113      &   2	         & ...        &       ...         &5 	 	      &   1           \\
 &UNK		 	  &   58      &   3          & 1          &       1           &6 	          &   3         \\
 \hline 
	\end{tabular}\\
Note: The classifiers and classes are presented in Column 1 and 2. Columns 3 lists the results of random forests Predictions in the work.
Columns 4-7 presents the results of comparison of  the Fan's predictions in  \cite{2019ApJ...879..107F}; 
the CKZ's  predictions in \cite{2022MNRAS.515.2215C} (where, Columns 6 lists the sources have CD) ; 
the Paliya's  Predictions in \cite{2021ApJS..253...46P};
the WIBRaLS2's results in \cite{2019ApJS..242....4D}
 for common objects. 
Where,  N is the number of the Fan's predictions, CKZ's predictions,  Paliya's  Predictions, and WIBRaLS2's results; 
M shows the number of random forests Predictions of the work in the cross-matching the FAN's predictions, CKZ's predictions,  
Paliya's  Predictions, and WIBRaLS2's results.
\end{table*}

Comparing the FBLs  with the TBLs  and the FSRQs reported in 4LAC (green open squares in right  panel in Figure  \ref{Fig_line}), we found that the ${\rm log}{F_{R}}$ with a median of -0.381 (or a mean of -0.434)
and $\Gamma_{\rm ph}$ with a median of 2.244 (or a mean of 2.265)
of the {FBLs} are both slightly larger than that of the TBLs 
with a median of -1.232  (or a mean of -1.259) and a median of 2.092 (or a mean of 2.089) respectively (see Table \ref{Tab_median}).
The ${\rm log}{F_{R}}$ of the FBLs  with a median of -0.381 (or a mean of -0.434)
and that of the FSRQs with a median of -0.404 (or a mean of -0.414)
 are overlapping and cannot be distinguished; 
However, the photon spectral index of  FBLs with a median of 2.244  (or a mean of 2.265)
 is slightly smaller than that of  FSRQs with a median of 2.450  (or a mean of 2.472).
The FBLs  are located in a regions where the ${\rm log}{F_{R}}$ is large relative to the {TBLs} and the the photon spectral index is small relative to the FSRQs.
These FBLs  may be intrinsically FSRQs with broad emission lines, which were mistaken for BL Lac-type sources due to their strong jet continuum swamping the broad emission lines and showing relatively small EW. 
When the continuum becomes weaker, the emission lines should exhibit a larger FSRQ-type EWs. 
The FBLs are the candidates 
for the transition from BL Lac to FSRQ (B-to-F transition), 
and the region (above line), where the FBLs are located in (see  Figure \ref{Fig_line}), can be referred to as the from BL Lac to FSRQ  transition region, called as B-to-F transition region (named as ``$B\rightarrow{F}$ zone”).

\section{Results comparison} \label{Results_comparison}
\vspace{3mm}

\subsection{Comparison with the predictions} \label{Results_comparison_Fan}
\vspace{3mm}

Cross-matching  the $C_9$ predictions (FBLs) and the predictions of  \cite{2019ApJ...879..107F}, 
of the 33 LSP BL Lac sources predicted as the possible FSRQ type sources by \cite{2019ApJ...879..107F}, 
there are 29 sources in our sample. Among the 29 possible FSRQ  type sources predicted by \cite{2019ApJ...879..107F}, the prediction results of 24 (24/29 $\simeq$ 82.76\%) sources  in the work are consistent with the results of  \cite{2019ApJ...879..107F}; 
However,  3 sources (4FGL J0428.6-3756; 4FGL J0811.4+0146 and  4FGL J0818.2+4222) are still no clear predictions;
and 2 sources (4FGL J0433.6+2905 and 4FGL J0738.1+1742) are predicted as true BL Lacs in our work (see Table \ref{Tab3} and  \ref{tab4_comp}).

\vspace{2mm}

In \cite{2021ApJS..253...46P},  the Compton dominance (CD; the ratio of the inverse Compton to synchrotron peak luminosities) for 1030   Fermi blazars are calculated. 
They found that the CD and accretion luminosity ($L_{\rm disk}$) in Eddington units ($L_{\rm disk}/L_{\rm Edd}$) are positively correlated, suggesting that the CD can be used to reveal the state of accretion in blazars and used to distinguish the classification of blazars. 
They suggest that blazars with CD $>$ 1 should be identified as FSRQs and CD $<$ 1 as BL Lac objects. 
There are 626 blazars with CD $>$ 1  in their sample. Cross-matching the $C_9$ prediction results  and  the 626 blazars  identified as FSRQs  in \cite{2021ApJS..253...46P}, we obtained 33 common sources. 
Among the 33 common objects, 22 (22/33 $\simeq$ 66.67\%) FSRQ candidates are consistent between our predictions and  \cite{2021ApJS..253...46P} predictions  (see Table \ref{Tab3}  and Table \ref{tab4_comp}).
5 TBLs (4FGL J0433.6+2905; 4FGL J1008.8-3139; 4FGL J1035.6+4409; 4FGL J1503.5+4759  and 4FGL J1942.8-3512) and 
6 UNKs  (4FGL J0032.4-2849 ; 4FGL J0428.6-3756; 4FGL J0811.4+0146; 4FGL J1331.2-1325; 4FGL J1647.5+4950  and 4FGL J1704.2+1234) were predicted in our work.

 \vspace{2mm}

The WIBRaLS2 catalog \citep{2019ApJS..242....4D} includes 9541 sources that
is an incremental version of the WISE Blazar-like Radio-Loud Sources (WIBRaLS) catalog 
(\citealt{2014ApJS..215...14D}).
Based on their WISE colors, 
these sources  are classified as 
3744 BL Lacs, 
5089 flat-spectrum radio quasars (BZQs), and
 708  mixed candidates.
Cross-matching  the $C_9$ predictions  and the 5089 BZQs  in \cite{2019ApJS..242....4D},
there are 37 BZQs in our prediction sample.
Among the 37 BZQs, 
the prediction results of 33 (33/37 $\simeq$ 89.19\%) sources in this work 
are consistent with the results of  \cite{2019ApJS..242....4D}; 
However,  3 sources 
(4FGL J0241.0-0505; 4FGL J0428.6-3756; and 4FGL J0818.2+4222) are still no clear predictions;
and 1 sources (4FGL J1135.1+3014) 
are predicted as true BL Lacs in this work (see Table \ref{Tab3} and  \ref{tab4_comp}).

 \vspace{2mm}

When cross-matching the $C_9$ predictions and the prediction results of \cite{2022MNRAS.515.2215C}, 39 of the 47 LSP BL Lac sources predicted as FSRQ are in our sample. 
The remaining 8 sources locate in the low Galactic latitudes ($|b|<10\deg$) 4LAC-DR2 catalog.
 Among the 39 sources, the prediction results of 38 sources  (38/39 $\simeq$ 97.44\%)  are consistent with our prediction results.
 There is only one source   (4FGL J2241.2+4120)  that does not have a clear prediction in the work (see Table \ref{Tab3}  and Table \ref{tab4_comp}). 
 In \cite{2022MNRAS.515.2215C},  they also check the CD of 39 
 sources based on SEDs fitting  using a quadratic polynomial.
For the 39 sources (see Columns 6 in Table \ref{tab4_comp}) reported the CD in  \cite{2022MNRAS.515.2215C},  31 sources are in our sample,  the other remaining 8 sources locate in the low Galactic latitudes ($|b|<10\deg$) 4LAC-DR2 catalog. 
There are  30 (30/31 $\simeq$ 96.77\%) sources' prediction are consistent with our predictions, in which 28 sources with CD $>$ 1; 
two sources with CD $<$ 1;
One source (4FGL J2241.2+4120 with CD $\simeq$ 3.689) that does not have a clear prediction in our work; 
As mentioned above, a comparison (the prediction results of approximately exceeding 
(24+38+30+22+33)/(29+39+31+33+37)  $\simeq$ 86.98\% are consistent)
of these predictions  suggests that our predictions may be promising.

\vspace{2mm}

\begin{figure*}
\centering
\includegraphics[width=17cm,height=7cm]{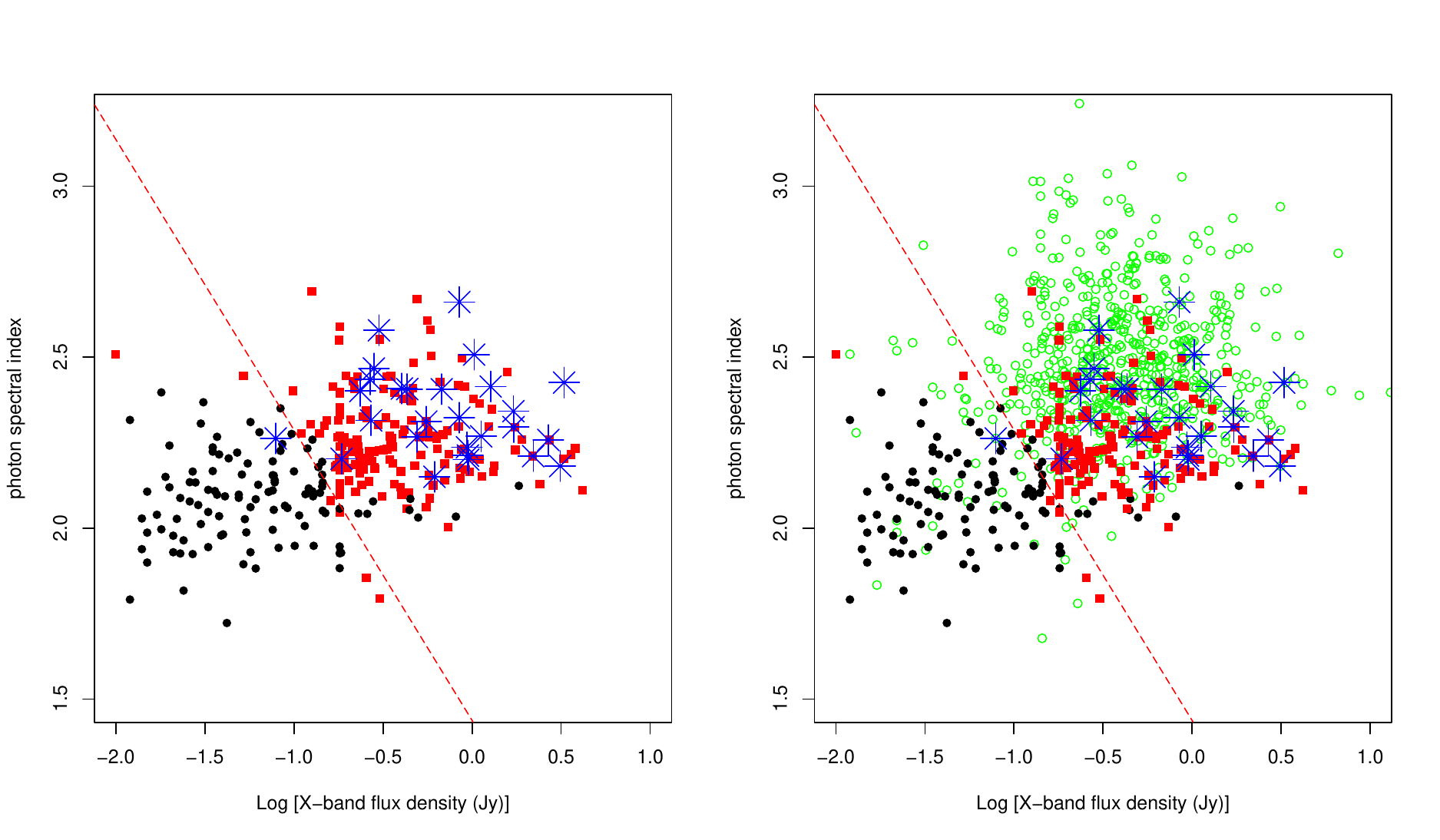}
\caption{Classification scatterplots for the Fermi $\gamma$-ray photon spectral index ($\Gamma_{\rm ph}$) and  the X-band VLBI radio flux (log $F_{R}$),
where the black filled circles, red solid squares, and  green empty circles indicate TBLs, FBLs and FSRQ respectively. 
And the blue stars represent  the LSP Changing-Look Blazars  reported  in \citealt{2022Univ....8..587F,2021Univ....7..372F}.
\label{Fig_LSP}}
\end{figure*}

\subsection{Comparison with the CLBs} \label{Results_comparison_CLB}
\vspace{3mm}

\begin{table}
	\centering
	\tiny
	\caption{The LSP CLBs reported in  Foschini et al. 2022.}
	\label{tab_CLB_Foschini}
	\begin{tabular}{cccccr} 
\hline\hline
{N} &{4FGL name}&{Optical class} & {SED class}  & {$\Gamma_{\rm ph}$} & {log$F_R$}\\
 \hline 
1	&	4FGL J0407.5$+$0741	&	bll	&	LSP	&	2.311 	&	-0.257 	\\
2	&	4FGL J1302.8$+$5748	&	bll	&	LSP	&	2.267 	&	-0.310 	\\
3	&	4FGL J1751.5$+$0938	&	bll	&	LSP	&	2.258 	&	0.428 	\\
4	&	4FGL J1800.6$+$7828	&	bll	&	LSP	&	2.210 	&	0.343 	\\
5	&	4FGL J2134.2$-$0154	&	bll	&	LSP	&	2.295 	&	0.233 	\\
 \hline 
6	&	4FGL J0217.8$+$0144	&	fsrq	&	LSP	&	2.236 	&	-0.029 	\\
7	&	4FGL J0449.1$+$1121	&	fsrq	&	LSP	&	2.507 	&	0.011 	\\
8	&	4FGL J0509.4$+$1012	&	fsrq	&	LSP	&	2.408 	&	-0.395 	\\
9	&	4FGL J0510.0$+$1800	&	fsrq	&	LSP	&	2.200 	&	-0.021 	\\
10	&	4FGL J0719.3$+$3307	&	fsrq	&	LSP	&	2.203 	&	-0.733 	\\
11	&	4FGL J0833.9$+$4223	&	fsrq	&	LSP	&	2.434 	&	-0.572 	\\
12	&	4FGL J1037.4$-$2933	&	fsrq	&	LSP	&	2.414 	&	0.104 	\\
13	&	4FGL J1043.2$+$2408	&	fsrq	&	LSP	&	2.323 	&	-0.072 	\\
14	&	4FGL J1124.0$+$2336	&	fsrq	&	LSP	&	2.406 	&	-0.365 	\\
15	&	4FGL J1224.9$+$2122	&	fsrq	&	LSP	&	2.342 	&	0.232 	\\
16	&   4FGL J1322.2$+$0842	&	fsrq	&	LSP	&	2.262 	&	-1.102 	\\
17	&	4FGL J1333.2$+$2725	&	fsrq	&	LSP	&	2.315 	&	-0.567 	\\
18	&	4FGL J1422.3$+$3223	&	fsrq	&	LSP	&	2.401 	&	-0.627 	\\
19	&	4FGL J1657.7$+$4808	&	fsrq	&	LSP	&	2.406 	&	-0.171 	\\
20	&	4FGL J1937.2$-$3958	&	fsrq	&	LSP	&	2.661 	&	-0.072 	\\
21	&	4FGL J2026.0$-$2845	&	fsrq	&	LSP	&	2.578 	&	-0.523 	\\
22	&	4FGL J2158.1$-$1501	&	fsrq	&	LSP	&	2.181 	&	0.495 	\\
23	&	4FGL J2212.0$+$2356	&	fsrq	&	LSP	&	2.212 	&	-0.024 	\\
24	&	4FGL J2225.7$-$0457	&	fsrq	&	LSP	&	2.425 	&	0.517 	\\
25	&	4FGL J2236.3$+$2828	&	fsrq	&	LSP	&	2.268 	&	0.051 	\\
26	&	4FGL J2345.2$-$1555	&	fsrq	&	LSP	&	2.150 	&	-0.211 	\\
27	&	4FGL J2349.4$+$0534	&	fsrq	&	LSP	&	2.467 	&	-0.551 	\\
 \hline 
	\end{tabular}\\
Note: The number of the records are presented in Column 1. 
Column 2 lists the source name of 4FGL.
The optical classes and the SED class reported in 4FGL are presented in Column 3 and Column 4, respectively, 
where “bll” indicates BL Lac and “fsrq” indicates FSRQ.
A simple horizontal line is used to distinguish  the FSRQs and BL lacs.
The $\gamma$-ray photon spectral index ($\Gamma_{\rm ph}$) and  the X-band VLBI radio flux (log$F_R$) 
are shown in Columns 5 and 6; respectively.
\end{table}

In \cite{2020ApJ...892..105A}, they found that the $\nu^{\rm S}_{\rm p}$ distributions are overlap between LSP BL Lacs  and FSRQs, 
and the gamma-ray photon spectral index ($\Gamma_{\rm ph}$) distributions are also very similar. 
They proposed that there is a possible region for objects that might be transitioning between FSRQs and BL Lacs.  
There are five of six such transitioning objects found by \cite{2014ApJ...797...19R} are the LSP subclass reported in 4LAC. 
In addition, \cite{2022ApJ...925...97P} also address a similar area, based on disk luminosity ($L_{\rm disk}$) in Eddington units ($L_{\rm disk}/L_{\rm Edd}$). 
They proposed that there is a region (called ``appareling zone”) with $2.00 \times 10^{-4}~\lesssim~L_{\rm disk}/L_{\rm Edd}~{\lesssim}~8.51 \times 10^{-3}$, 
where,  some sources (that are perhaps changing-look blazars) may be a transition between BL Lacs and FSRQs. 
And they found five confirmed changing-look sources reside in the ``appareling zone”.

\vspace{2mm}

In the work, a B-to-F transition zone (transition from BL Lac to FSRQ) is suggested by comparing the prediction results: TBLs and FBLs. In order to  test the effectiveness of the B-to-F zone. 
Where, whether
some LSP BL Lacs (e.g., FBLs) are located in the ``$B\rightarrow{F}$ zone” that are the most likely Candidates of Changing-Look Blazar.
We check a LSP CLBs sample that was collected by \cite{2022Univ....8..587F,2021Univ....7..372F}.
All of which are compiled into an online (Transition)  Changing-Look Blazars Catalog (TCLB Catalog\footnote{\url{https://github.com/ksj7924/CLBCat}}, \url{https://github.com/ksj7924/CLBCat}) (Kang et al., in preparation) presented in https://github.com/ for easy communication. 
In \cite{2022Univ....8..587F,2021Univ....7..372F}, they compiled a gamma-ray jetted  AGN sample based on the 4FGL catalog.
They reported 34 Changing-Look AGNs, 32 of them are labeled as blazars (24 FSRQs, 7 BL Lacs, and 1 BCU) in 4FGL catalog, 
based on 
a significant change in optical spectral lines  (disappearance and reappearance) in different observation epochs
reported in the previous literature 
(see \citealt{2022Univ....8..587F,2021Univ....7..372F} for more details and references therein). 
Among the them, there are 27 LSP CLBs (22 LSP FSRQ type and 5 LSP BL Lac type) in our sample and 
which are listed in Table \ref{tab_CLB_Foschini}.
A simple horizontal line is used to distinguish  the LSP FSRQs and LSP BL lacs labeled in 4FGL.

\begin{figure*}
\centering
\includegraphics[width=17cm,height=7cm]{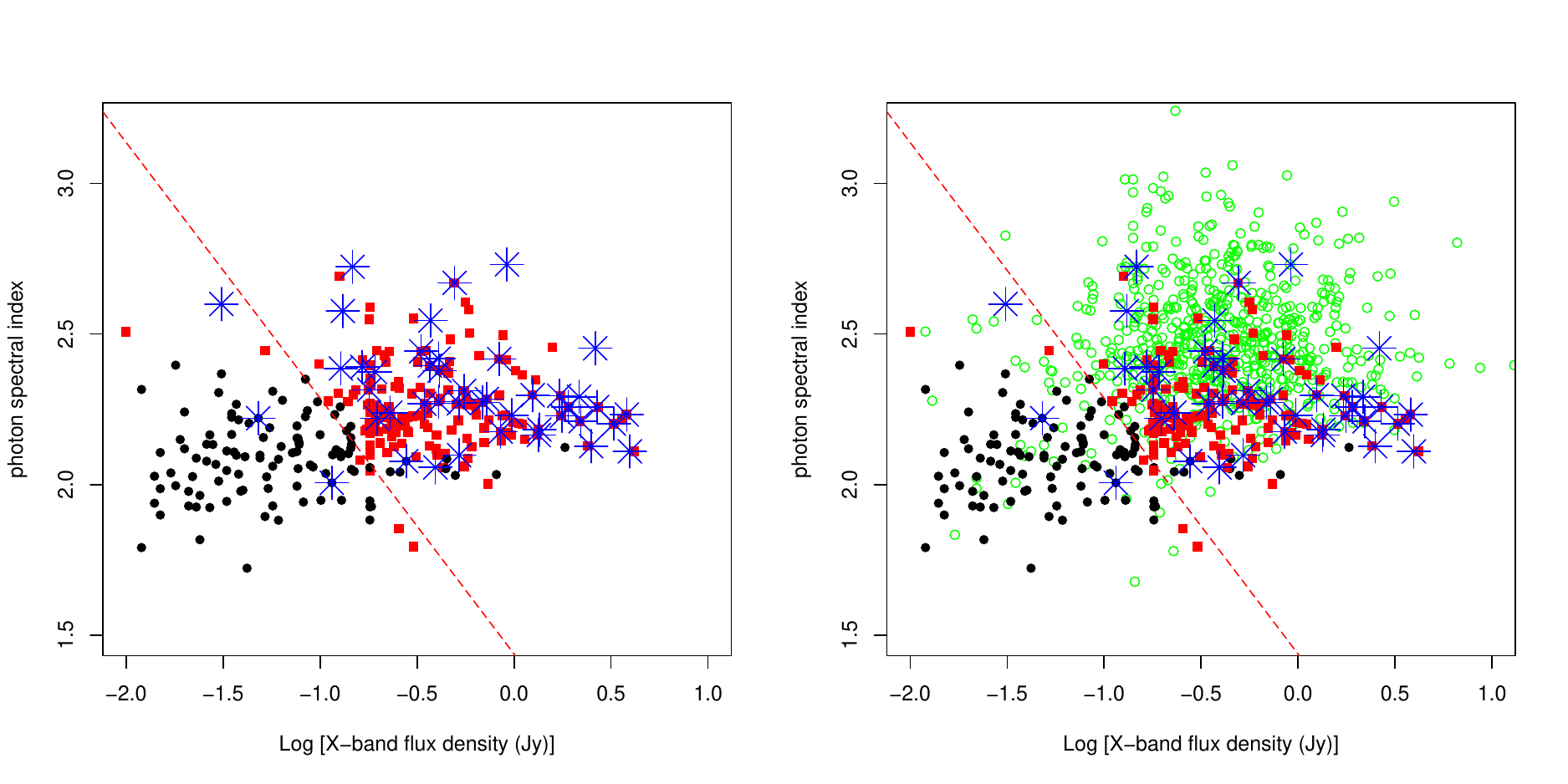}
\caption{Classification scatterplots for the Fermi $\gamma$-ray photon spectral index ($\Gamma_{\rm ph}$) and  the X-band VLBI radio flux (log $F_X$),
where the black filled circles, red solid squares, and  green empty circles indicate TBLs, FBLs and FSRQs respectively. 
And the blue stars represent  the LSP BL Lac  type
Changing-Look  Blazars  reported in Xiao et al. (2022a).}
\label{Fig_LSP_fan}
\end{figure*}

\vspace{2mm}

For ease of comparison, the classification scatterplots for the Fermi $\gamma$-ray photon spectral index ($\Gamma_{\rm ph}$) and  the X-band VLBI radio flux (log$F_{R}$) are plotted in Figure \ref{Fig_LSP}, where the black filled circles, red solid squares, and  green empty circles indicate TBLs, FBLs predicted in this work and FSRQs labeled in 4FGL, respectively. 
And the blue stars represent  the LSP CLBs (22 LSP FSRQ type and 5 LSP BL Lac type sources) are reported in \citealt{2022Univ....8..587F,2021Univ....7..372F}.
The left panels represent the correlation between $\Gamma_{\rm ph}$ and log$F_{\rm R}$ of the TBLs and FBLs predicted in this work and CLBs identified in \citealt{2022Univ....8..587F,2021Univ....7..372F}. In the right panels, the FSRQs labeled in 4FGL are also added (see, green circle).

\vspace{2mm}

In Figure \ref{Fig_LSP}, for the entire LSP CLBs,
we note that most of the LSP CLBs (26 of 27  sources, $\simeq$ 96.29\%) are located in the B-to-F transition zone.
Only one source, 4FGL J1322.2+0842 with $\Gamma_{\rm ph} = 2.262$, and log$F_{R} = -1.102 $ labeled as LSP FSRQ in 4FGL catalog, that do not locate in the B-to-F transition zone.
Where, there are 21 FSRQ type CLBs, which is that the LSP BL Lacs  that may have transitioned from BL Lacs to FSRQs.
The results suggest the B-to-F transition region is valid for LSP BL Lac. 
Where some LSP BL Lacs (e.g., FBLs) are located in the ``$B\rightarrow{F}$ zone” that are the most likely Candidates of Changing-Look Blazars.

\vspace{2mm}

In addition, recently, in \cite{2022ApJ...936..146X}, 
they reported  52 new Changing-look blazars  (10 FSRQs and 42 BL Lacs) 
based on the EW classification of their optical spectra.
There are 45 LSP CLBs  in our sample and are listed in Table \ref{tab_CLB_fan}, 
which includes 10 LSP FSRQ type CLBs and 35 LSP BL Lac  type CLBs.
A simple horizontal line is used to distinguish  the FSRQs and BL lacs.
All the 35 LSP BL Lacs that are in our  forecast sample.
Among the 35 LSP BL Lacs, 
the prediction results of 29 (29/35 $\simeq$ 82.86\%)  sources are consistent with our prediction results; 
3 TBLs (4FGL J0433.6$+$2905, 4FGL J0654.7$+$4246, and  4FGL J1503.5$+$4759) and 
3 UNKs (4FGL J0428.6$-$3756, 4FGL J1331.2$-$1325, 4FGL J1647.5$+$4950) were predicted in our work.

\vspace{2mm}

In Figure \ref{Fig_LSP_fan}, these CLBs reported in \cite{2022ApJ...936..146X} also were plotted.
For the 10 LSP FSRQ type CLBs, that are all located in the B-to-F transition region.
For the 35 LSP BL Lac  type CLBs (see Table \ref{tab_CLB_fan}),
we note that except for one source (4FGL J1058.0+4305) without radio data (see Table \ref{tab_CLB_fan});
three source (4FGL J0654.7+4246 and 4FGL J1503.5+4759 are evaluated as TBLs, and 4FGL J1331.2-1325 is evaluated as UNKs) is not located in the B-to-F transition region;
31 of them are all located in the B-to-F transition region.
For the entire 45 LSP CLBs, 
there are 41 (41/45 $\simeq$ 91.11\%) LSP CLBs  (31 BL Lacs and 10 FSRQs) are located in the B-to-F transition region.
These results also further indicate that the B-to-F transition region is valid for the LSP BL Lac subclass.
Where some LSP BL Lacs (e.g., FBLs) are located in the ``$B\rightarrow{F}$ zone” that are the most likely Candidates of Changing-Look Blazars.

\begin{table}
	\centering
	\tiny
	\caption{The Changing-Look  blazars reported in Xiao et al. 2022a.}
	\label{tab_CLB_fan}
	\begin{tabular}{ccccccc} 
\hline\hline
{N} &{4FGL name}&{Optical class} & {SED class}  & {$\Gamma_{\rm ph}$} & {log$F_R$}\\
 \hline 
1	&	4FGL J0006.3$-$0620	&	bll	&	LSP	&	2.128 	&	0.397 	\\
2	&	4FGL J0203.7$+$3042	&	bll	&	LSP	&	2.239 	&	-0.638 	\\
3	&	4FGL J0209.9$+$7229	&	bll	&	LSP	&	2.275 	&	-0.389 	\\
4	&	4FGL J0238.6$+$1637	&	bll	&	LSP	&	2.165 	&	0.127 	\\
5	&	4FGL J0334.2$-$4008	&	bll	&	LSP	&	2.183 	&	0.131 	\\
6	&	4FGL J0407.5$+$0741	&	bll	&	LSP	&	2.311 	&	-0.257 	\\
7	&	4FGL J0428.6$-$3756	&	bll	&	LSP	&	2.098 	&	-0.283 	\\
8	&	4FGL J0433.6$+$2905	&	bll	&	LSP	&	2.078 	&	-0.556 	\\
9	&	4FGL J0438.9$-$4521	&	bll	&	LSP	&	2.417 	&	-0.077 	\\
10	&	4FGL J0516.7$-$6207	&	bll	&	LSP	&	2.176 	&	-0.069 	\\
11	&	4FGL J0538.8$-$4405	&	bll	&	LSP	&	2.111 	&	0.597 	\\
12	&	4FGL J0629.3$-$1959	&	bll	&	LSP	&	2.231 	&	-0.012 	\\
13	&	4FGL J0654.7$+$4246	&	bll	&	LSP	&	2.006 	&	-0.939 	\\
14	&	4FGL J0710.9$+$4733	&	bll	&	LSP	&	2.670 	&	-0.306 	\\
15	&	4FGL J0831.8$+$0429	&	bll	&	LSP	&	2.273 	&	-0.162 	\\
16	&	4FGL J0832.4$+$4912	&	bll	&	LSP	&	2.378 	&	-0.384 	\\
17	&	4FGL J1001.1$+$2911	&	bll	&	LSP	&	2.269 	&	-0.462 	\\
18	&	4FGL J1058.0$+$4305	&	bll	&	LSP	&	2.346 	&	…	\\
19	&	4FGL J1058.4$+$0133	&	bll	&	LSP	&	2.233 	&	0.581 	\\
20	&	4FGL J1058.6$-$8003	&	bll	&	LSP	&	2.258 	&	0.280 	\\
21	&	4FGL J1147.0$-$3812	&	bll	&	LSP	&	2.229 	&	0.247 	\\
22	&	4FGL J1250.6$+$0217	&	bll	&	LSP	&	2.057 	&	-0.406 	\\
23	&	4FGL J1331.2$-$1325	&	bll	&	LSP	&	2.600 	&	-1.509 	\\
24	&	4FGL J1503.5$+$4759	&	bll	&	LSP	&	2.221 	&	-1.319 	\\
25	&	4FGL J1647.5$+$4950	&	bll	&	LSP	&	2.386 	&	-0.893 	\\
26	&	4FGL J1751.5$+$0938	&	bll	&	LSP	&	2.258 	&	0.428 	\\
27	&	4FGL J1800.6$+$7828	&	bll	&	LSP	&	2.210 	&	0.343 	\\
28	&	4FGL J1806.8$+$6949	&	bll	&	LSP	&	2.297 	&	0.096 	\\
29	&	4FGL J1954.6$-$1122	&	bll	&	LSP	&	2.443 	&	-0.479 	\\
30	&	4FGL J2134.2$-$0154	&	bll	&	LSP	&	2.295 	&	0.233 	\\
31	&	4FGL J2152.5$+$1737	&	bll	&	LSP	&	2.270 	&	-0.284 	\\
32	&	4FGL J2202.7$+$4216	&	bll	&	LSP	&	2.202 	&	0.515 	\\
33	&	4FGL J2216.9$+$2421	&	bll	&	LSP	&	2.284 	&	-0.140 	\\
34	&	4FGL J2315.6$-$5018	&	bll	&	LSP	&	2.395 	&	-0.434 	\\
35	&	4FGL J2357.4$-$0152	&	bll	&	LSP	&	2.315 	&	-0.747 	\\
 \hline 
36	&	4FGL J0102.8$+$5824	&	fsrq	&	LSP	&	2.292 	&	0.335 	\\
37	&	4FGL J0337.8$-$1157	&	fsrq	&	LSP	&	2.724 	&	-0.833 	\\
38	&	4FGL J0347.0$+$4844	&	fsrq	&	LSP	&	2.577 	&	-0.883 	\\
39	&	4FGL J0521.3$-$1734	&	fsrq	&	LSP	&	2.374 	&	-0.717 	\\
40	&	4FGL J0539.6$+$1432	&	fsrq	&	LSP	&	2.545 	&	-0.431 	\\
41	&	4FGL J0539.9$-$2839	&	fsrq	&	LSP	&	2.731 	&	-0.037 	\\
42	&	4FGL J0601.1$-$7035	&	fsrq	&	LSP	&	2.418 	&	-0.389 	\\
43	&	4FGL J1816.9$-$4942	&	fsrq	&	LSP	&	2.391 	&	-0.783 	\\
44	&	4FGL J2015.5$+$3710	&	fsrq	&	LSP	&	2.453 	&	0.418 	\\
45	&	4FGL J2121.0$+$1901	&	fsrq	&	LSP	&	2.220 	&	-0.706 	\\
 \hline 
	\end{tabular}\\
Note: The number of the records are presented in Column 1. 
Column 2 lists the source name of 4FGL.
The optical classes and the SED class reported in 4FGL are presented in Column 3 and Column 4, respectively, 
where “bll” indicates BL Lac and “fsrq” indicates FSRQ.
A simple horizontal line is used to distinguish  the FSRQs and BL lacs.
The $\gamma$-ray photon spectral index ($\Gamma_{\rm ph}$) and  the X-band VLBI radio flux (log$F_R$) 
are shown in Columns 5 and 6; respectively.
\end{table}

\section{Discussion  and Conclusion } \label{Discuss_Conclusion}
\vspace{3mm}

A portion of BL Lacs, which showed the observational characteristics of FSRQ type sources, could be potential FSRQs, vice versa.
The peculiar rare transition phenomenon that have found between BL Lacs and FSRQs (e.g., EW become larger or smaller) has been addressed by some authors.
Which is common addressed by some possible scenarios in the previous literature
(e.g., see \citealt{2021AJ....161..196P} for the related discussions and references therein).
For instance, the broad lines (EW) of some transition sources may be swamped by the 
strong  (beamed) jet continuum variability (e.g., 
\citealt{1995ApJ...452L...5V};
\citealt{2012MNRAS.420.2899G};
\citealt{2014ApJ...797...19R};
\citealt{2019RNAAS...3...92P}), or 
jet bulk Lorentz factor variability (e.g., 
\citealt{2009A&A...496..423B}); 
Or some transition sources with weak radiative cooling, the broad lines are overwhelmed by the non-thermal continuum  (e.g., \citealt{2012MNRAS.425.1371G}).
In addition,  some strong broad lines of the FSRQ type source are missed due to with a high redshift (e.g., $z > 0.7$, \citealt{2015MNRAS.449.3517D}), the one of the strongest $H{\alpha}$ line falls outside the optical window, caused the misclassification.
Also, several observational effects (e.g., signal-to- noise ratio, and spectral resolution, etc.)  may also affected the optical classification (see \citealt{2021AJ....161..196P} for the related discussions).
In-depth research is of great significance to deepen the understanding of the origin of transition (e.g., CLB type) sources, the accretion state transition of supermassive black holes; jet  particle acceleration process; and black hole-galaxy co-evolution, etc (e.g., \citealt{2014ApJ...797...19R}; \citealt{2021AAS...23740807M}).

\vspace{2mm}

Based on the 4LAC, 4FGL and RCF catalog, we constructed a sample containing 1680 Fermi sources with known EW-based (optical) classifications (FSRQs and BL Lacs) and SED-based classifications (LSP, ISP, and HSP). Which includes  651 FSRQs and 1029 BL Lacs, that are divided into 960 LSP, 334 ISP and 386 HSP sources. Where, 1352 blazars with 651 FSRQs and 701 ISP and HSP BL Lacs are viewed as the training and validation samples;  All 328 LSP BL Lacs are viewed as a forecast sample. Approximately 4/5 of 1352 blazars are randomly (random seed = 123) assigned to the training sample, and the remaining ones (e.g., approximately 1/5) are considered as the validation sample. Here, the training sample include 1082 blazars with 528 FSRQs and  554 HSP (and LSP) BL Lacs, and the validation sample has  270 blazars with 123 FSRQs and  147 ISP, or HSP BL Lacs. Based on the D $>$ {0.3} in the two sample K-S test and Gini $>$ 2.1 in random forests algorithm for all parameters with valid observations, there are 23 parameters selected in the work for the 1680 sources. Using the the random forests algorithm with the default settings for the random forests classification functions  ($randomForest()$ in R code) to the selected sample (the training, validation and forecast samples), the 8388607 different combinations for the selected 23 parameters are calculated. There are 178  OPCs (the optimal parameters combinations) with a maximum accuracy (accuracy$\simeq$0.9889) are obtained, where, 1,  5, 14, 35, 52, 39, 28,  2, or 2 combinations  of  5,  6,  7,  8,  9, 10, 11, 12, or 13 parameters (see Table \ref{tab_numb}), respectively. We select nine combinations,  one combination in the combinations with 5, 6, 7, 8, 9, 10, 11, 12, or 13 parameters, respectively. Combined the classification results from the nine optimal combinations of parameters, 113 TBLs and 157 possible FBLs are predicted;  however, 58 remain without a clear prediction; for 328 LSP BL Lacs reported in the high Galactic latitudes ($|b|>10\deg$) 4LAC-DR2 catalog.

\vspace{2mm}

In section \ref{Results_comparison}, the prediction results between our ($C_9$) predictions of this work using random forests algorithm and 
the Fan's predictions \citep{2019ApJ...879..107F}, CKZ's predictions \citep{2022MNRAS.515.2215C}, Paliya's  Predictions \citep{2021ApJS..253...46P},
or  WIBRaLS2 catalog's results \citep{2019ApJS..242....4D} 
are compared, respectively.
Among the common objects through cross-matching, the prediction results of most of the sources are consistent (see Table \ref{tab4_comp}). 
It suggests that our predictions are robust and effective.
However, we should note that among the 1680 selected sample sources, there are 167 source X-band flux data missing 
in the training and validation sample and 30 source X-band flux data missing in the forecast sample.
In the random forests algorithm, the missing data is filled with the median using the $na.roughfix()$ function of the random forests algorithm.
Based on a set of combined parameters, we test that the predictions of missing data with mean filling 
are basically consistent with that of with the median filling; 
We also tested, delete the source of missing data, the prediction results are basically the same.
However, the prediction accuracy decreased slightly (accuracy $\simeq $0.9728), which basically had no effect on our main conclusions.
Also, there are a total of 178 optimal parameters combinations (OPC) with a maximum accuracy and we select only 9 of these OPCs to combine and construct our prediction results ($C_9$ predictions, see Table \ref{tab_resultA} and \ref{Tab3}).
We also tried crossing more parameter combinations, the predictions FBLs and TBLs had slightly fewer sources, and UNK sources had slightly more sources. From various combinations, we select only a part of them, which have a little effect on the final prediction result.
In addition, some sources without SED classes were removed, possibly with selection effects. In the random forest calculation, the $randomForest()$ function using the default setting, and the random factor is set as: seed=123, also slightly affect the forecast results (see \citealt{2019ApJ...872..189K} for some related discussion), also need to note.

\vspace{2mm}

Moreover, we also should note that the misclassified BL Lacs are singled out only from the LSP BL Lacs and only based the observational properties: 22 Fermi-specific parameters plus a radio flux from the RFC in the work.
For the ISP/HSP BL Lacs as possible FBLs that should have relatively high powers and peak frequency ($\nu^{\rm S}_{\rm p}$) values of the synchrotron bump
(e.g., \citealt{1998ApJ...506..621G};
\citealt{2012MNRAS.420.2899G};
\citealt{2017ApJ...834...41K};
\citealt{2019MNRAS.484L.104P} and references therein)
and other parameters (e.g., Eddington ratios, equivalent to the accretion rates and powers, etc),
which are not considered (dismissed) in the work, may be further addressed in future work.

\vspace{2mm}

In addition,  we also should note that 
the analysis in this work is purely based in gamma-rays and radio-frequencies data 
and the sources were evaluated using a random forest algorithm. 
We only evaluated that some BL Lacs showed the observational characteristics of FSRQ type sources
could be potential FSRQs.
However, for their true classifications, an optical spectroscopic follow up would be needed to prove it.
We should know that 
blazars usually display large amplitude and rapid variability across the electromagnetic spectrum.
The EW value measured from optical spectra is common based on a single-epoch classification.
Which may be affected by the variability of blazars.
For instance, the broad lines (EW) may be swamped by the 
strong  (beamed) jet continuum variability (e.g., 
\citealt{1995ApJ...452L...5V};
\citealt{2012MNRAS.420.2899G};
\citealt{2014ApJ...797...19R};
\citealt{2019RNAAS...3...92P}).
However, the 4FGL catalog is based on 10 years of observations of Fermi-LAT data 
and its parameter's values are calculated from central tendency values.
Which may be less affected by the variability of blazars and more reflective of its intrinsic characteristics than the value of EW.
Therefore, the issues in the context may be more reasonable.
We expect further research to verify whether the hypothesis is reasonable or not.
However, its essential classification still needs to be confirmed by optical spectra.

\vspace{2mm}

Comparing the predictions: between TBLs  and FBLs,  we find that the {FBLs} show a clear separation for TBLs in  the $\Gamma_{\rm ph}$-${\rm log}{F_{R}}$ plane, which can use a simple phenomenological critical line (see Equation \ref{equ_line} and Figure \ref{Fig_line}) to roughly separate these two subclasses, where, the {FBLs} are located in higher areas. 
Checking the LSP CLBs, there are 26 of 27 LSP CLBs are located in the transition zone.
Therefore, we propose a B-to-F transition zone named ``$B\rightarrow{F}$ zone” where the transition from BL Lac to FSRQ will occur for LSP BL Lac. Where, some LSP BL Lacs  (e.g., FBLs) are located in the “B → F zone” that are the most likely Candidates of Changing-Look Blazar.

\vspace{2mm}
We note that the EW of the changing-look blazar B2 1420+32 is close to 5{\AA}, and floats  around 5{\AA}, sometimes the EW is greater than 5{\AA}, sometimes less than 5{\AA}, during between FSRQ and BL Lac state transitions \citep{2021ApJ...913..146M}.
For the FBLs that are the possible Candidates of Changing-Look Blazar found in the work, whether the EWs of them are also close to 5{\AA},
we will do further verification next.

\begin{figure*}
\centering
\includegraphics[width=17cm,height=6cm]{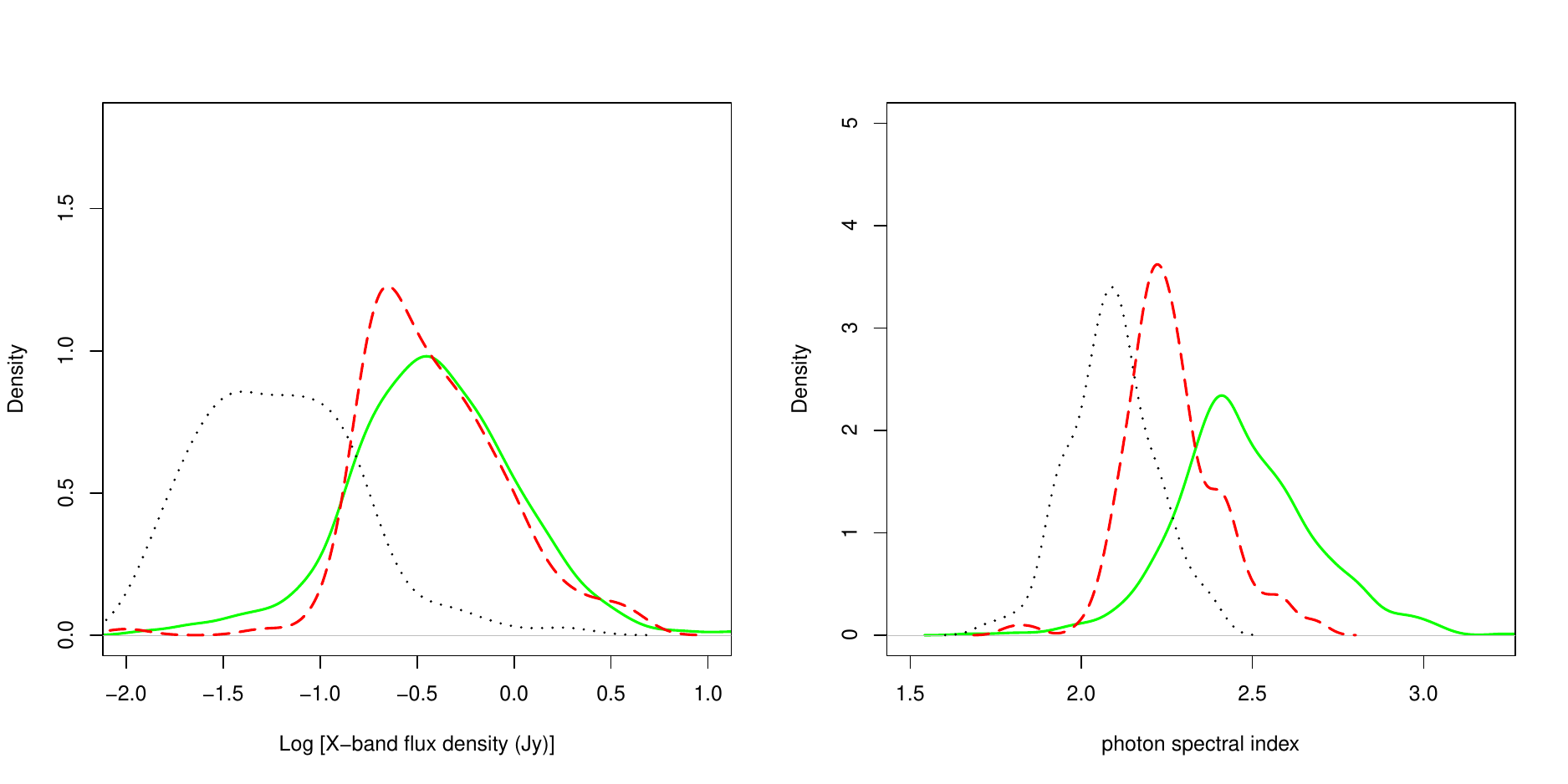}
\caption{the density distribution for the photon spectral index ($\Gamma_{\rm ph}$, right  panel) and  the X-band flux (log $F_{R}$, left  panel), 
where the black dashed lines, 
red long-dashed lines,
and 
green solid lines
indicate TBLs,  FBLs , and FSRQs respectively.
 \label{Fig_density}}
\end{figure*}

\begin{figure}
\centering
\includegraphics[width=8cm,height=6cm]{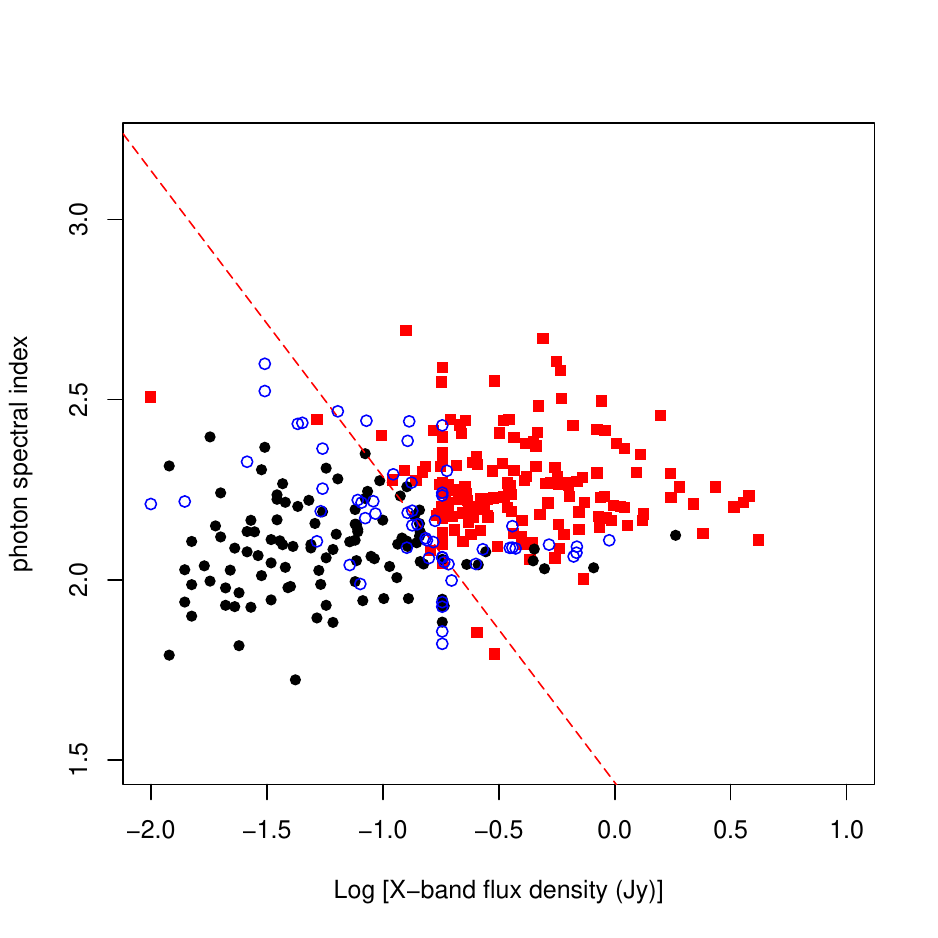}
\caption{Scatterplots for the photon spectral index ($\Gamma_{\rm ph}$) and  the X-band flux (log $F_{R}$),
where the black filled circles, red solid squares, and  blue empty circles indicate TBLs, FBLs and UNKs respectively. \label{Fig_unk}}
\end{figure}

\vspace{2mm}

Moreover, we should note that we only address the transition from LSP BL Lacs to FSRQs in the work and only suggest a ``$B\rightarrow{F}$ zone”. Using the B-to-F transition zone, the LSP FBLs can be well identified from LSP BL Lacs.
We also note that all the LSP BL Lac type CLBs are located in the B-to-F transition region (see Figure \ref{Fig_LSP} in Section \ref{Results_comparison}). 
Almost all the LSP FSRQ type CLBs are located in the B-to-F transition region, only one source  
(4FGL J1322.2$+$0842  with $\Gamma_{\rm ph}$  = 2.262 and  ${\rm log}{F_{R}}$ = -1.102) are not in.
But, the source is very close to the critical line.
Which imply that these LSP FSRQ type CLBs may be some LSP BL Lacs that have transitioned, may be BL Lacs before the transition (e.g., see \citealt{2021Univ....7..372F,2022Univ....8..587F},  some FSRQs, featureless or weaker lines in previous literature). This also further indicates that the B-to-F transition zone can effectively screen out those potential FBLs (e.g., Changing-Look type blazars)
from LSP BL Lacs, which verifies the validity/effectiveness of the B-to-F transition zone.
In addition, 
vice versa, a transition from FSRQs to BL Lacs (F-to-B transition) is also possible. 
The possible candidates for the F-to-B transition that cannot be addressed only based on this work. 
Which needs to be further addressed in the future.
When the F-to-B transition candidate sources are also screened, 
the F-to-B transition region can be effectively demarcated.
Also, the complete CLBs (B$\rightleftarrows$F) transition region (B-to-F transition, Vice versa) may be proposed.
And the role of CLB in the blazar sequence, its evolution (e.g., blazar redshift evolution), origin and other issues can be further discussed/studied (\citealt{2014ApJ...797...19R}).
Also, the ``$B\rightarrow F$ zone” is obtained only based on the two parameters of the ${\rm log}{F_{R}}$ and $\Gamma_{\rm ph}$.
We also tried to check all possible combinations of the 23 parameters, but there was no other better findings.
For other parameters (or multi-dimensional parameter space) whether there are similar regions, or better discrimination, further research is needed.

 \vspace{2mm}

In addition, we also should note that the density distribution of  the $\Gamma_{\rm ph}$ (right  panel in Figure \ref{Fig_density}) and  the log $F_{R}$ (left  panel) for  the TBLs,  FBLs, and FSRQs show a different distributions.
The ${\rm log}{F_{R}}$ of the TBLs and that of the FSRQ are clearly distinguished,
where the D = 0.858  and p $<$ 1E-16 for the two-sample K-S test,
however, the ${\rm log}{F_{R}}$ of the FBLs and that of the FSRQ are overlapping, 
where the D = 0.096  and p = 0.226 for the two-sample K-S test,
and cannot be distinguished.
Which seems to indicate that {FBLs} and FSRQs have similar radio jet properties.
The $\Gamma_{\rm ph}$ of the FBLs  (with a median of 2.244, see Table \ref{Tab_median})
located between that of the TBLs  (with a median of 2.092)
and the FSRQs (with a median of 2.450), 
that are slightly larger than that of the TBLs, and smaller than that of the FSRQs. 
Which may imply that the FBLs are in an intermediate transition state between the TBLs and the FSRQs.
Whether it is similar to or related to the transition of the accretion mode of the central engine (Kang et al., in preparation), 
or other related physical changes, requires further in-depth study. It will be of great significance to understand the evolution of blazars and other blazar jets physics.


We also note that there are 58 UNK sources without a clear prediction (see Table \ref{Tab3}).
They are scattered on both sides of the critical line (see Figure \ref{Fig_unk}).
The boundary between FBLs and TBLs is simply distinguished employed a straight line, 
which is a bit too simple and seems a bit unreasonable.
Complex separation boundaries (or multi-dimensional parameter space) may be more realistic and effective, 
which needed to be further addressed.
The LSP CLB source: 4FGL J1322.2$+$0842 is not located in the B-to-F transition zone, but, the source is very close to the critical line.  
Combined with the distribution characteristics of 58 UNK sources,
it seems imply that there are more complex separation boundaries.
Whether the assumption is reasonable and whether it exists requires further research.


Although there are still many deficiencies in our work, 
our work may be still effective in diagnosing the  possible of FBLs from LSP BL Lacs. 
For extremely rare CLB sources, our work will greatly enrich the sample of CLB sources, 
which would influence the study of different properties between BL Lacs and FSRQs, 
especially regarding the role of CLB sources in the evolution of blazar sequences (Kang et al., in preparation), or their redshift evolution, etc., and provide abundant samples. 
It would also provide potentially valid target sources for the discovery of additional CLB sources and  for subsequent confirmation of CLB sources, especially future large spectroscopic or photometric surveys. 
This issue will continue to be addressed in future work.


\section*{Acknowledgements}

We thank the anonymous editor and referee for very constructive and helpful comments and suggestions, which greatly helped us to improve our paper. 
This work is partially supported by the National Natural Science Foundation of China 
(Grant Nos. 12163002, U1931203, and 12233007, 12363003), 
the National SKA Program of China (Grant No. 2022SKA0120101),
the Guizhou Provincial Basic Research Program (Natural Science, Grant No. QKHJC-ZK[2024]ZD-000),
and
the Discipline-Team of Liupanshui Normal University (Grant No. LPSSY2023XKTD11).

\section*{Data Availability}

The data underlying this article will be shared on reasonable request to the corresponding author.


\bibliographystyle{mnras}
\bibliography{mnras} 


\bsp	
\label{lastpage}
\end{document}